\documentclass[aps,pra,reprint,superscriptaddress]{revtex4-2}

\usepackage{booktabs}
\usepackage[colorlinks=true,allcolors=blue]{hyperref}

\usepackage[utf8]{inputenc}
\usepackage[english]{babel}
\usepackage[T1]{fontenc}
\usepackage{lmodern}
\usepackage{amsmath,amsfonts,amssymb,amsthm,bm,times,dcolumn}
\usepackage{microtype}
\usepackage{gensymb} % for example \degree symbol can be used!
\usepackage{physics}
\usepackage{color}
\usepackage{soul}
\usepackage[normalem]{ulem}
\usepackage{graphicx,color}

\begin{document}

%\title{A depth-fidelity frontier for collision-model Dicke-state preparation, its circuit cost, and measured fidelities on a  superconducting processor}

\title{Data and code for collision-model Dicke-state preparation: depth-fidelity frontiers, circuit costs, and superconducting-processor measurements}

\author{Duc-Kha Vu}
\affiliation{Department of Electrical and Computer Engineering, Saint Louis University, St Louis, Missouri, 63103, USA}

\author{Minh Tam Nguyen}
\affiliation{KT One, 4302 Shire Court, Tampa, FL, 33613, United States}
\affiliation{University of South Florida, 4202 E Fowler Ave, Tampa, FL 33620, United States}

\author{\c{S}ahin K. \"{O}zdemir}	
\affiliation{Department of Electrical and Computer Engineering, Saint Louis University, St Louis, Missouri, 63103, USA}

\author{\"{O}zg\"{u}r E. M\"{u}stecapl{\i}o\u{g}lu}
\affiliation{Department of Physics, Ko\c{c} University, Sar{\i}yer, \.Istanbul, 34450, Türkiye}
\affiliation{TÜB\.ITAK Research Institute for Fundamental Sciences (TBAE), 41470 Gebze, Türkiye}

\author{Fatih Ozaydin}
\email{mansursah@gmail.com}
\affiliation{Institute for International Strategy and Emerging Technologies, Tokyo International University, 4-42-31 Higashi-Ikebukuro, Toshima-ku, Tokyo 170-0013, Japan}
%\affiliation{Data Science \& AI Major, Tokyo International University, 4-42-31 Higashi-Ikebukuro, Toshima-ku, Tokyo 170-0013, Japan}
\affiliation{Nanoelectronics Research Center, Kosuyolu Mah., Lambaci Sok., Kosuyolu Sit., No:9E/3  Kadikoy, Istanbul, T\"urkiye}

\date{\today}

\begin{abstract}
	Dicke states are multipartite entangled states in which a fixed number of quantum excitations is coherently shared among many qubits.
	Originally introduced in the context of cooperative emission and superradiance, they are now important resources for quantum sensing, networking, and collective quantum phenomena.
	Preparing prescribed Dicke states with high fidelity, however, remains challenging, particularly as the system size and excitation number increase.
	Here we present an open dataset and accompanying code for preparing Dicke states using a collision-based quantum protocol.
	The dataset covers systems from five to fourteen qubits over a broad range of excitation numbers and records how the best-found noiseless preparation fidelity changes with circuit depth.
	It also provides circuit-resource estimates and experimental measurements for selected states on the 54-qubit IQM Emerald superconducting processor.
	The accompanying code reproduces the processed data and validation checks, providing a reusable benchmark for studying the trade-off between state-preparation fidelity, circuit cost, and hardware noise.
\end{abstract}


\maketitle

\section{Introduction}
\label{sec:intro}

Multipartite entanglement is an indispensable resource for quantum networking, distributed sensing, and measurement-based quantum protocols. Within the broader class of symmetric multiqubit states, Dicke states hold a particularly prominent position, owing to their rich entanglement structure and broad relevance to quantum information processing~\cite{dicke1954coherence}. Originally introduced to describe cooperative
emission~\cite{rehler1971superradiance,gross1982superradiance}, Dicke states provide a natural framework to study collective radiance in cavity and
waveguide QED~\cite{amsuss2011cavity,sheremet2023waveguide}, and they also serve as a versatile resource for metrological enhancement in symmetric
subspaces~\cite{hotter2024combining,saleem2024achieving,gubaydullin2026quantum},
entanglement detection through collective observables~\cite{lohof2023signatures, lucke2014detecting}, and multipartite quantum networking tasks~\cite{chiuri2012experimental, cheng2020realizing}. A recent superconducting-qubit magnetometry study provides a relevant example: in a phase-estimation magnetometer based on superconducting fluxonium sensors subject to relaxation and dephasing, a comparison of Greenberger–Horne–Zeilinger (GHZ) and Dicke probes revealed that Dicke-state probes give up some peak local sensitivity in exchange for a substantially broader dynamical range and enhanced robustness to noise and bias errors~\cite{gubaydullin2026quantum}. The single-excitation member of the family, the $W$ state, is the canonical example whose pairwise entanglement survives particle loss~\cite{dur2001multipartite,neven2018entanglement} and decoherence~\cite{sen2003multiqubit}, while the states with higher excitation numbers support greater entanglement depth and a richer correlation structure~\cite{chen2016entanglement}, but are more demanding to prepare. The entanglement and coherence structure of the
family continues to attract attention~\cite{bhattacharyya2025entanglement,bhattacharyya2026super}.

The preparation of Dicke states is a longstanding challenge, even in the single-excitation case.
Since $W$ states cannot be enlarged deterministically by local operations alone, early photonic approaches focused on probabilistic local expansion and transformation.
Linear-optical schemes were proposed to enlarge an existing $W$ state by acting on only one of its photons~\cite{tashima2008elementary,tashima2009local}, and related work demonstrated both the local conversion of two EPR pairs into a three-photon $W$ state~\cite{tashima2009localPRL} and the experimental expansion of photonic $W$ states to larger sizes~\cite{tashima2010demonstration}.
A complementary line of research developed probabilistic fusion schemes in which smaller $W$ states are combined into larger entangled states, with Fredkin-gate, cross-Kerr, and related constructions aimed at improving success probabilities and resource requirements~\cite{bugu2013enhancing, yesilyurt2013optical,ozaydin2014fusing,zang2015generating,li2016generating}.
Other optical state-engineering approaches have also considered the distillation of imperfect $W$ states, including metamaterial-assisted schemes~\cite{al2015quantum}.
Deterministic expansion strategies instead expand a small $W$ state with separable ancillary qubits~\cite{yesilyurt2016deterministic, zang2016deterministic,ozaydin2021deterministic}.
Additional platform-specific approaches include preparation through Pauli spin blockade~\cite{bugu2020preparing} and linear-optical protocols~\cite{kim2020efficient}.

Beyond the single-excitation sector the constructions become substantially more demanding. Proposals for arbitrary
excitation number include circuit families with explicit gate-count
bounds~\cite{chakraborty2014efficient,bartschi2019deterministic,mukherjee2020preparing,aktar2022divide,yu2024efficient},
global control of an Ising-coupled register~\cite{stojanovic2023dicke},
phase estimation in a spin ensemble~\cite{wang2021preparing}, adiabatic
and counterdiabatic
driving~\cite{linington2008robust,opatrny2016counterdiabatic,carrasco2024dicke},
engineered dissipation~\cite{zhu2025dissipation}, unitary
transformations between Dicke states~\cite{kobayashi2014universal}, and
expansion of a four-qubit Dicke state to a five-qubit one under
restricted qubit access~\cite{thapa2025expanding}. 
Most recently, Wang et al. proposed a deterministic
cavity-mediated adiabatic protocol for preparing a prescribed Dicke
state along the symmetric Dicke ladder, using a programmable detuning,
a coherent transverse drive, and an effective collective-spin
interaction~\cite{wang2026deterministic}.
Nearly all of these approaches share the same paradigm: gates are treated as the resource and the number of gates used as the cost of preparation, while fidelity in the absence of a noise model serves as the figure of merit.

On current hardware, gate count and noiseless fidelity alone are insufficient to determine experimental performance. 
Two-qubit error rates can vary by more than an order of magnitude across the couplings of a single device, and finite coherence times constrain the total duration of a
circuit~\cite{preskill2018quantum,arute2019quantum}. Thus, circuit duration directly affects the fidelity attainable at measurement. Although error mitigation can recover expectation values, its sampling cost grows with circuit
depth~\cite{temme2017error,endo2021hybrid}, sharpening rather than
removing the constraint. Optimizing only noiseless fidelity does not penalize circuit depth and may therefore select unnecessarily deep operating points. 

Collision models provide a natural setting in which to examine this trade-off. In a
collision, or repeated-interaction, model a system is driven by a
sequence of short unitary encounters with ancillary
units~\cite{ziman2002diluting,ciccarello2022quantum,campbell2021collision};
such models can simulate any multipartite Markovian
dynamics~\cite{cattaneo2021collision}, have been compiled into digital
circuits with explicit resource
counts~\cite{erbanni2023simulating,garg2025simulating}, and have been
run on noisy processors to reproduce collective dissipative
effects~\cite{cattaneo2023quantum}. Used constructively rather than as a
model of an environment, a mobile shuttle ancilla mediating
excitation-preserving partial-swap collisions between otherwise
uncoupled registers generates genuine multipartite
entanglement~\cite{ccakmak2019robust}. Ref.~\cite{vu2026intelligent} treats the collision strengths as design parameters and optimizes them so that the registers converge
deterministically on a target Dicke state, removing the projective
measurement on the shuttle and extending the reachable family beyond the
single-excitation sector. Because the number of collisions is the free
variable of that construction and the objective is noiseless
fidelity, the operating points it reports extend to many collision rounds.

For the same protocol and target states, this release captures
the quantities that the choice of operating point has to trade
against each other: the fidelity attainable at each circuit depth, the
circuit cost of each of those operating points, and the fidelity and the circuit cost returned by a superconducting processor for a subset of these operating points. The three quantities are reported together because the measured fidelity should be interpreted relative to both the noiseless fidelity of the same circuit and its gate count. 
The trade-off is evident in the measurements: the shallowest arms retain around three-quarters of their noiseless fidelity, while the deepest approach the value returned by a fully decohered circuit.

The release is organized into three layers with the corresponding files listed in Table~\ref{tab:inventory}. The layers are linked: L1 defines the operating points, L2 assigns circuit costs to those points, and L3 reports hardware measurements for a selected subset.

\textbf{L1, the noiseless frontier.} For each of the 31 target states $D_N^K$ and each with a maximum allowed
depth $m=0,\ldots,200$, we report the highest fidelity found at or
below that depth, together with the collision strengths and the
sub-round at which that fidelity is attained. Its size is simply the product of these two numbers: 31 states $\times$
201 depths $= 6,231$ cells, all fully populated. The search
underlying each cell is released too, in the form of three arrays per state, each holding one value for every candidate at every depth: 256 candidates $\times$ 201 depths $= 51,456$ values per state per array, which is 1,595,136 values across all 31 states.

\textbf{L2, circuit cost.} One row for each of those 6,231 cells reports the number of
partial-swap interactions, the corresponding all-to-all
two-qubit gate count, and two routing-inclusive estimates.
The interaction counts are exact when the required collision-strength
information is available; the remaining 544 rows carry conservative
upper bounds, as detailed in Sec.~\ref{sec:l2}.

\textbf{L3, hardware.} Ten circuit arms across four target states were executed on IQM Emerald, with 88,656 shots collected. The release includes the raw counts, submitted execution plans, calibration snapshots, and processed fidelities.

\subsection{Relation to the companion paper}
\label{sec:relation}

The protocol, and a first optimization of its collision strengths, are
reported in the companion paper~\cite{vu2026intelligent}. The two papers address the same protocol but serve different purposes. Ref.~\cite{vu2026intelligent} interprets selected operating points and their implications for shallow circuits, whereas the present article documents the complete release, its generation, and its validation. The hardware measurements common to both works are included here as
part of the complete reusable record, together with the raw counts,
metadata, and verification materials described below.
What this release adds to the material behind Ref.~\cite{vu2026intelligent}:

\begin{enumerate}
\item \emph{The whole depth axis rather than a stopping point.}
  Ref.~\cite{vu2026intelligent} reports a converged operating point per state.
 This release reports the best-found fidelity for every maximum depth
 from $m=0$ to $m=200$ sub-rounds, giving 6,231 frontier cells across
 the 31 target states.
\item \emph{Every candidate, not only the winners.} The three
  arrays per state hold all 1,595,136 candidate evaluations,
  which determine the selection rule's output at any depth and the
  pre-phase overlap of Eq.~(\ref{eq:invariance}).
\item \emph{A stated selection rule.} The phase layer is optimized
  at every depth and the winner is chosen on post-phase fidelity over
  both candidate and depth. Fixing the depth before optimizing the
  phase layer is not equivalent: it changes the result in
  8 of the 31 states, by up to
  +0.068 in fidelity.
\item \emph{A second, independent search.} A separate optimization
  over $m=0,\ldots,49$ sub-rounds is released alongside the deeper one,
  because neither dominates the other, together with the rule that
  combines them.
\item \emph{A circuit-cost layer}, absent from Ref.~\cite{vu2026intelligent}, provides partial-swap counts, corresponding all-to-all
two-qubit gate counts, and fitted routing overheads for every cell.
The release explicitly identifies the 544 cells for which the
interaction counts are conservative upper bounds rather than exact
values.
\item \emph{The hardware record}, released in full: raw measurement counts, submitted execution plans, the corresponding calibration snapshot, and all verification scripts, rather than summary values alone.
\end{enumerate}

\subsection{Intended reuse}
\label{sec:reuse}

The three layers were assembled to be usable separately as well as together. Together, they support five principal reuse cases.

\emph{Benchmarking a better search.}
Each L1 cell is a best-found value produced by the released searches
and the selection rule of Sec.~\ref{sec:l1methods}, rather than a
certified global optimum.
At each evaluated depth, each search compares 256 refined
collision-strength candidates.
For each state, the starting point, refined pair, and resulting loss are
provided for all 256 candidates, alongside the box constraints and the selection rule of
Sec.~\ref{sec:l1methods}. This allows an alternative optimizer to be run on the same problem and compared cell by cell, rather than only at a single
converged point. The files \texttt{\_raw.npy} and \texttt{\_final.npy} record the results before and after optimization of the phase layer, respectively.
This separation allows a user to test a different optimizer for either the collision strengths or the phase layer while keeping the other stage unchanged.

\emph{Choosing an operating point under a gate budget.} L1 and L2 are joined
on (state, \texttt{budget}), turning the frontier into a fidelity-versus-cost curve for each target state. A user with a fixed
two-qubit gate budget can invert this curve to identify the deepest circuit
that fits within the budget and the fidelity it reaches, rather than adopting a converged operating point and discovering its cost only afterwards. 
For cells supplied by the deeper search, the recorded collision
strengths allow the partial-swap and all-to-all two-qubit gate counts
to be determined exactly.
The 544 cells supplied by the shallower search lack the
collision-strength information required for an exact cost calculation,
so their reported costs are conservative upper bounds.
These cells are identified by their provenance in
\texttt{frontier\_source.csv}.
The routing-inclusive columns are estimates for all rows, as discussed
in Sec.~\ref{sec:limitations}.

\emph{Re-estimating fidelity from the same shots.} The raw counts for
all ten arms are provided together with the per-qubit assignment-error matrices from the calibration snapshot, allowing the data to be reprocessed with estimators other than that of
Sec.~\ref{sec:fidelity}. Full permutationally invariant tomography, maximum-likelihood
reconstruction, a different readout unfolding, or a bootstrap over shots
instead of over measurement settings can each be run on the deposited
counts; the last of these directly addresses the three-degree-of-freedom limitation discussed in Sec.~\ref{sec:limitations}.

\textit{Scoring a noise model against measured depth dependence.}
The depth series contains measurements at five circuit depths for each state.
At each depth, we report the excitation-sector population $P_K$, defined as the
probability of finding the system in the subspace with the target number $K$ of
excitations.
These measurements are provided together with the corresponding calibration
snapshot.
Section~\ref{sec:design} fits two variants to it,
one that adjusts the two-qubit gate duration and one that inflates every
calibrated gate error by a single factor. Both fits are deposited, so that alternative models can be scored against the same points. The layout experiment provides an empirical measure of the sensitivity
of the measured conditional fidelity to the choice of physical-qubit
layout on this device.

\emph{A test fixture for reimplementations.} Equation~(\ref{eq:invariance})
holds at all 1,595,136 stored candidate-depth pairs to a largest
relative deviation of $6\times10^{-8}$. A violation in an independent implementation would indicate an error in either excitation conservation or the implemented collision structure, so the invariance serves as a unit test and
not only as an observation about these arrays;
\texttt{code/pi\_raw\_invariant.py} evaluates it.

Section~\ref{sec:limitations} specifies the limitations of the dataset, and Sec.~\ref{sec:usage} defines the counting conventions required when reporting individual cells.

\section{Protocol and PARAMETERIZATION}
\label{sec:protocol}

This section provides only what is needed to use the released
parameters.

The target is the Dicke state $D_N^K$, the equal-weight superposition
of all $\binom{N}{K}$ computational basis states of $N$ qubits with
$K$ excitations. $K$ of $N$ qubits act as shuttle ancillas and
carry the excitations in. The remaining $N-K$ form two registers, $R$
and $S$, of sizes $\lfloor (N-K)/2 \rfloor$ and $\lceil (N-K)/2
\rceil$. \emph{The two registers are never coupled to each other.}
Every interaction in the protocol is either between an ancilla and a
register site, or between neighboring sites inside the same register;
no interaction joins $R$ to $S$. In Figure~\ref{fig:protocol}, we place the
ancillas between the registers so that this is visible.

Each interaction is a partial swap (pSWAP), a two-qubit unitary set by two
angles: $\gamma_{\rm sh}$ for an ancilla-register collision and
$\gamma_{\rm in}$ for a within-register collision. After the
collisions, a layer of single-qubit phase rotations is applied to all
$N$ qubits.

A collision operation with strength $\gamma$ is modeled by the operation
\begin{equation}
	U(\gamma) = \cos\gamma \, I + i \sin\gamma \, \mathrm{SWAP}
	= e^{\, i \gamma \, \mathrm{SWAP}} ,
	\label{eq:collision}
\end{equation}
where $I$ is the two-qubit identity operator and $\mathrm{SWAP}^2 = I$. 
%	Since $\mathrm{SWAP} = (I + X \otimes X +
%	Y \otimes Y + Z \otimes Z)/2$, the collision equals $\exp[i\gamma
%	(X \otimes X + Y \otimes Y + Z \otimes Z)/2]$ up to a global
%	phase, which is isotropic Heisenberg exchange left on for a time
%	proportional to $\gamma$. It is therefore a fractional SWAP: a full
%	SWAP at $\gamma = \pi/2$, and the identity up to a phase at
%	$\gamma = \pi$. 
Using $\mathrm{SWAP}
=
\frac{1}{2}
\left(
I + X\otimes X + Y\otimes Y + Z\otimes Z
\right),$
where $X$, $Y$, and $Z$ are the Pauli operators, Eq.~(1) can be written as $U(\gamma) = e^{i\gamma/2}
\exp\!\left[
\frac{i\gamma}{2}
\left(
X\otimes X + Y\otimes Y + Z\otimes Z
\right)
\right]$.
The factor $e^{i\gamma/2}$ is a global phase and therefore has no
observable effect. The remaining unitary has the form generated by an isotropic Heisenberg
exchange interaction, with $\gamma$ determining the interaction strength
(or, equivalently, the interaction time for a fixed coupling strength).
Thus, $U(\gamma)$ represents a fractional SWAP operation.
For $\gamma=\pi/2$, it becomes a full SWAP up to a global phase, whereas
for $\gamma=\pi$ it reduces to the identity up to a global phase.
	It is not a controlled gate, and on a
	superconducting device it is not a native operation either, so it
	has to be compiled. Figure~\ref{fig:protocol}(c) gives the
	compilation: three CZ interleaved with single-qubit rotations, for
	any $\gamma$ that is not a multiple of $\pi$. That is also why
	panel (a) marks a collision with
	the SWAP symbol, a cross at each end of the vertical line, rather
	than with the filled dots of a controlled gate.

\subsection{Depth, and how it is counted}
\label{sec:depth}

A \emph{sub-round} is one ancilla completing its sweep over both
registers: it visits $r_1, s_1, r_2, s_2, \dots$ in turn, and when
$\gamma_{\rm in} \neq 0$ each visited site then collides with its
neighbors inside its own register. A \emph{full round} is $K$
sub-rounds, one per ancilla. Throughout this release, depth is measured in sub-rounds, denoted by $m$, with $m=0,...,200$. A full
round corresponds to $m = K$. Thus the deepest point in the release is
200 sub-rounds, which corresponds to $200/K$ full rounds, yielding a different number of full rounds for each state.

\begin{figure*}[t!]
  \includegraphics[width=2\columnwidth]{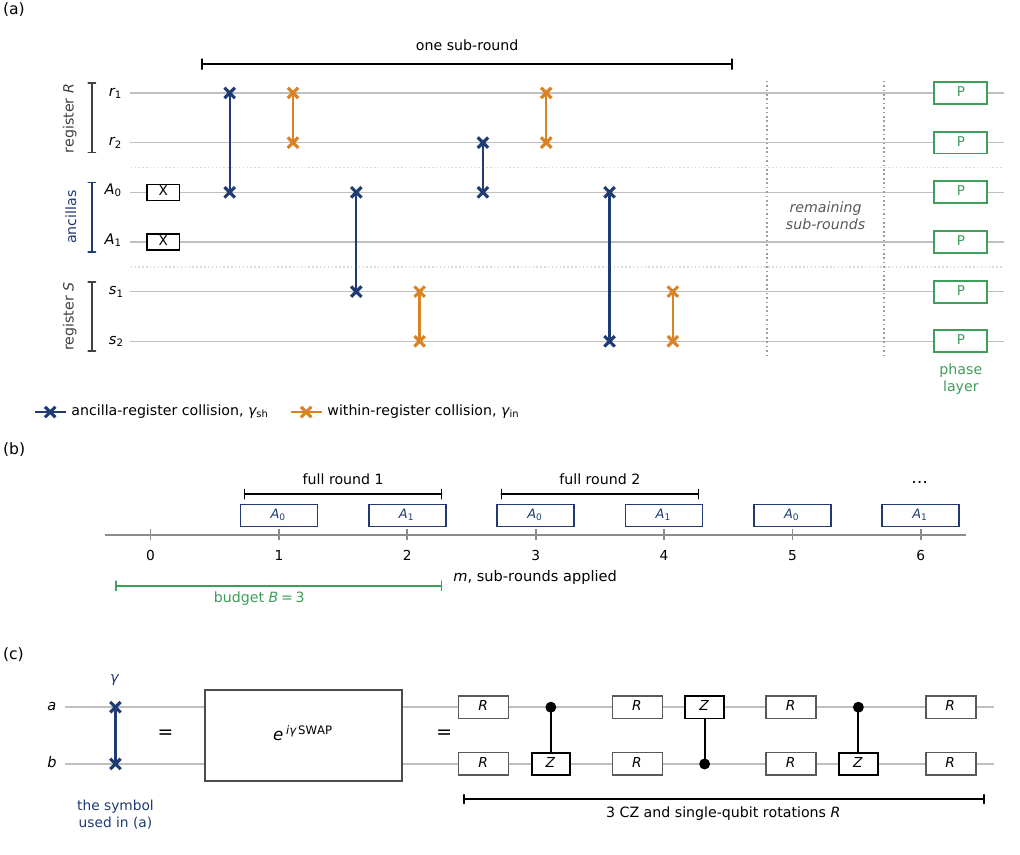}
\caption{\label{fig:protocol}
	The protocol and depth convention illustrated for $D_{6}^{2}$.
	(a) One sub-round.
	For $D_{6}^{2}$, $K=2$ shuttle ancillas, $A_0$ and $A_1$, lie between
	two two-qubit registers, $R$ and $S$; no interaction directly connects
	the two registers.
	Ancilla $A_0$ sequentially interacts with each register site through
	ancilla--register collisions (blue), and each visited site then
	interacts with its neighbors within the same register through
	within-register collisions (orange).
	For a two-site register, the single internal edge is used twice per
	sub-round.
	A layer of single-qubit phase rotations $P$ closes the circuit.
	(b) Depth $m$ counts sub-rounds, with one full round every $K$
	sub-rounds.
	The 1-based budget index $B$ used in
	\texttt{frontier\_canonical.csv} covers all depths $m\leq B-1$;
	in the example shown, $B=3$ covers $m=0,1,2$.
	(c) The collision operator decomposed into elementary quantum gates.
	The crossed vertical line denotes the collision operator of
	Eq.~(\ref{eq:collision}), an exchange interaction corresponding to a
	fractional SWAP rather than a controlled gate.
	In the CZ decomposition shown, a nontrivial collision is implemented
	using three CZ gates interleaved with the indicated single-qubit gates.
}
\end{figure*}

The \texttt{budget} column is 1-based, $B = m+1$, because its first row is
the circuit with no collisions applied. The row $B$ reports the best
fidelity over all depths up to and including $m = B-1$, and the
\texttt{subround} column records which depth achieved it. One full
round therefore appears in row $B = K+1$. Figure
\ref{fig:protocol}(b) shows the two indices side by side.

The number of collisions in a sub-round is set by the size of the registers,
not by $N$ alone, and it is not one within-register collision per
ancilla-register collision: a site at the end of a register has one
neighbor, an interior site has two, and a register of one qubit has
none. For $D_{6}^{2}$, drawn in Figure~\ref{fig:protocol}, a
sub-round is 8 collisions,
4 of them ancilla-register. For
$D_{8}^{3}$ it is 11 and
5. Setting $\gamma_{\rm in} = 0$
removes the within-register collisions from the circuit altogether.
Consequently, \(\gamma_{\rm in}\) determines both the interaction strength and whether within-register collisions are present in the circuit.
	
\section{Data records}
\label{sec:records}

The released materials are distributed as a single archive.
Table~\ref{tab:inventory} summarizes the directory structure and the
contents of each data layer.
Detailed definitions of all table columns are provided in
\texttt{DATA\_DICTIONARY.md}.

In addition to the directories listed in Table~\ref{tab:inventory},
the archive root contains 11 files for documentation, metadata,
reproducibility, and integrity checking.
These include \texttt{README.md} and three additional documentation
files, the license file, and the machine-readable metadata files
\texttt{CITATION.cff}, \texttt{codemeta.json}, and
\texttt{datapackage.json}.
The root directory also contains the pinned
\texttt{requirements.txt}, the \texttt{reproduce.py} entry point,
and \texttt{CHECKSUMS.sha256}, which provides checksums for all files
in the archive.

\begin{table*}[tb]
  \caption{Deposit contents, by directory.}
  \label{tab:inventory}
  \begin{ruledtabular}
  \footnotesize
  \begin{tabular}{ll r p{0.55\textwidth}}
    layer & path & files & contents \\
    \colrule
    L1 & \texttt{data/L1\_frontier/} & 195 & the canonical frontier over 6,231 cells, its provenance, both source campaigns, the per-state winners and the full optimization surfaces \\
    L2 & \texttt{data/L2\_gate\_costs/} & 2 & 6,231 rows of partial-swap counts, the all-to-all two-qubit floor and routing-inclusive estimates; the five-regime ladder \\
    L3 & \texttt{data/raw/} & 11 & raw counts, the submitted plans, the device calibration snapshot, the depth series and the layout experiment \\
    L3 & \texttt{data/processed/} & 16 & fidelities and error bars for 10 arms, the arm comparisons, the noise-model refit and every verification record \\
    -- & \texttt{data/parameters/} & 3 & collision strengths as a standalone table, and the published operating points used as comparison arms \\
    -- & \texttt{code/} & 25 & the analysis and verification chain, with a pinned environment \\
    -- & \texttt{figures/} & 15 & the figures of this paper as PDF and PNG, each with the exact series it plots as JSON \\
  \end{tabular}
  \end{ruledtabular}
\end{table*}

\subsection{L1: the noiseless frontier}
\label{sec:l1}

\texttt{frontier\_canonical.csv} contains one column for each of the
31 target states and 201 rows indexed by the budget variable
$B=1,\ldots,201$, giving 6,231 populated cells in total.
The target states span $N=5,\ldots,14$ qubits and excitation numbers
$K=2,\ldots,5$.
For a given state and budget $B$, the corresponding entry gives the
highest fidelity found over all circuit depths
$m\leq B-1$, obtained from exact statevector simulation without a
noise model.
The file \texttt{frontier\_source.csv} records which search contributes
each cell.
The file \texttt{frontier\_summary.json} provides, for each target
state, the fidelity after one full round, the best fidelity found over
all depths, the difference between these values, and the shallowest
depth at which 99\% of the best-found fidelity is reached.

The released dataset contains 31 target states.
Four additional states were computed but excluded:
$D_{12}^{6}$, $D_{13}^{6}$, $D_{14}^{6}$, and $D_{14}^{7}$.
These four states were covered only by the shallower search and not by
the deeper search.
For the shallower search, the per-candidate collision strengths needed
to determine the exact circuit cost are not available in the deposit.
Consequently, all L2 cost entries for these four states would have been
upper bounds rather than exact counts.
They were therefore excluded from the released state set.
The exclusion list and its rationale are encoded in
\texttt{code/pi\_states.py}, and the analysis scripts apply this list
consistently.

\begin{figure*}[tb]
  \includegraphics[width=\textwidth]{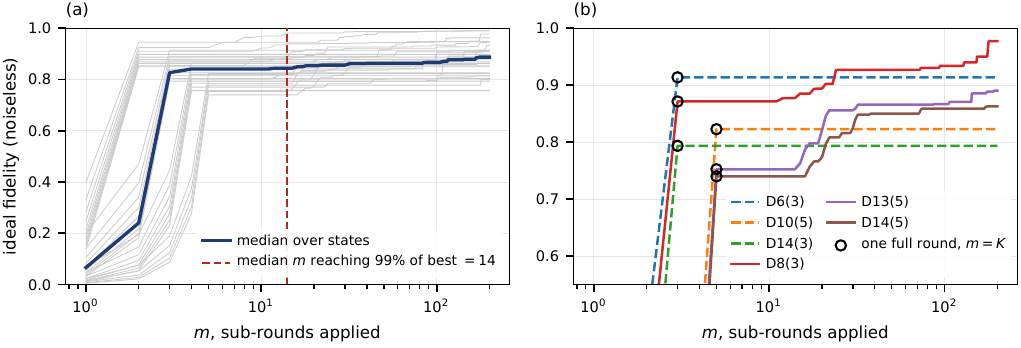}
%\caption{\label{fig:frontier}
%	The L1 fidelity frontier.
%	(a) Best-found noiseless fidelity as a function of the maximum allowed
%	depth $m$, measured in sub-rounds, for all 31 target states.
%	The across-state median and the median depth required to reach 99\% of
%	each state's best-found fidelity are also shown.
%	(b) Six representative target states.
%	Dashed curves denote states for which one full round already gives the
%	best-found fidelity within the explored depth range; open circles mark
%	one full round, $m=K$.
%	Depth $m=0$, corresponding to the circuit before any collision is
%	applied, is not shown.}
	\caption{\label{fig:frontier}
		The L1 fidelity frontier.
		(a) Best-found noiseless fidelity versus the maximum allowed depth
		$m$, measured in sub-rounds, for all 31 target states.
		Gray curves show the individual target states, and the solid curve
		shows the across-state median.
		The vertical dashed line marks the median depth, $m=14$, required to
		reach 99\% of each state's best-found fidelity.
		(b) Six representative target states.
		Dashed curves indicate states for which the best-found fidelity is
		already attained after one full round, $m=K$; open circles mark the
		one-full-round points.
		Depth $m=0$, corresponding to a circuit with no collisions, is omitted
		for clarity.
	}
\end{figure*}

Across the 31 released states, the median difference between the
best-found fidelity and the fidelity after one full round is $0.0223$.
The largest difference is $0.1382$ for $D_{13}^{5}$.
Figure~\ref{fig:frontier}(a) shows the depth dependence for all target
states together with the across-state median.
Figure~\ref{fig:frontier}(b) shows six representative examples.
For 14 of the 31 states, one full round already gives the highest
fidelity found within the explored depth range.
The remaining states show further improvement at larger depths.
Across all states, the median depth required to reach 99\% of the
best-found fidelity is 14 sub-rounds, with values ranging from 2 to
192 sub-rounds.

The underlying search results are provided as three arrays for each
target state.
Each array contains $256\times201=51{,}456$ values, corresponding to
256 candidates evaluated over 201 depths.
Across the 31 released states, this gives 1,595,136 values per array.
The file \texttt{\_raw.npy} contains the values before optimization of
the phase layer, whereas \texttt{\_final.npy} contains the corresponding
values after phase optimization.
The file \texttt{\_absbound.npy} contains the associated modulus bound.
Applying the selection rule described in
Sec.~\ref{sec:l1methods} to \texttt{\_final.npy} reproduces the
published frontier cell for each state.
The pre-phase overlap appearing in Eq.~(\ref{eq:invariance}) is
evaluated using \texttt{\_raw.npy}.

\subsection{L2: circuit cost}
\label{sec:l2}

The file \texttt{gatecounts\_frontier.csv} contains one row for each
of the 6,231 L1 cells.
For each cell, it reports circuit cost at three levels of increasing
hardware specificity.

\begin{figure*}[t!]
  \includegraphics[width=2\columnwidth]{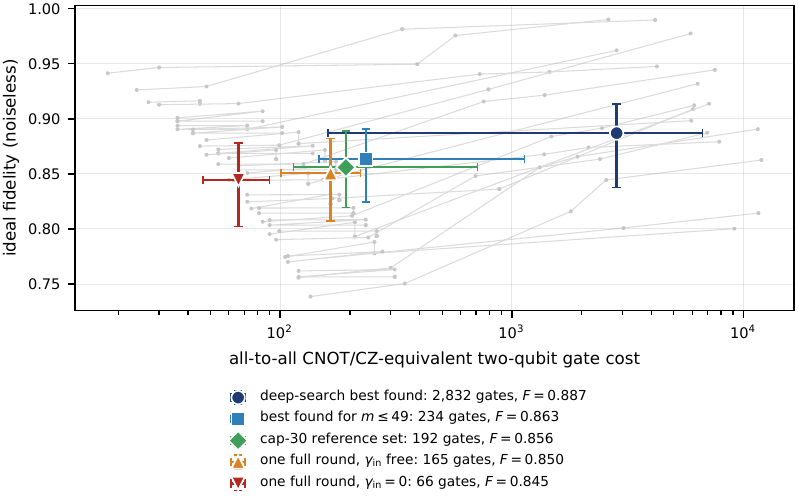}
%	\caption{\label{fig:cost}
%		Noiseless fidelity versus the all-to-all CNOT/CZ-equivalent two-qubit
%		gate cost for the five operating regimes of
%		Table~\ref{tab:regimes}.
%		Each gray line connects the five operating regimes for one target
%		state.
%		Colored markers show the across-state medians, with horizontal and
%		vertical bars spanning the corresponding interquartile ranges.
%		The two-qubit gate cost is $3\times\mathrm{pSWAP}$ as defined in
%		Sec.~\ref{sec:l2}; routing overhead is not included.}
\caption{\label{fig:cost}
	Noiseless fidelity versus the all-to-all CNOT/CZ-equivalent two-qubit
	gate cost for the five operating regimes summarized in
	Table~\ref{tab:regimes}.
	Each gray line connects the five operating regimes for one target
	state.
	Colored markers show the across-state medians, with horizontal and
	vertical bars spanning the corresponding interquartile ranges.
	Here pSWAP denotes the number of partial-swap interactions, and the
	two-qubit gate cost is $3\times\mathrm{pSWAP}$ as defined in
	Sec.~\ref{sec:l2}; routing overhead is not included.
}
\end{figure*}

\begin{enumerate}	
	\item \texttt{pswap} gives the number of partial-swap (pSWAP) interactions
	in the circuit.
	This quantity counts protocol interactions rather than hardware
	gates.
	For rows with complete collision-strength information, it is
	obtained exactly from a closed-form expression that depends on
	$N$, $K$, the circuit depth, and whether
	$\gamma_{\rm in}=0$.
	The expression was verified against every stored transpiled
	circuit.
	The 544 rows for which the exact interaction count cannot be
	determined are discussed below.
	
%	\item \texttt{n2q\_all\_to\_all} gives the corresponding two-qubit
%	gate count for an idealized device with all-to-all connectivity.
%	For a nontrivial partial swap, this count is
%	$3\times\texttt{pswap}$.
%	The factor of three follows from the structure of the unitary in
%	Eq.~(\ref{eq:collision}), rather than from an empirical overhead.
%	Its three canonical (Weyl) coordinates are equal and are nonzero
%	unless $\gamma$ is a multiple of $\pi$.
%	A two-qubit unitary with three nonzero canonical coordinates
%	cannot be realized using fewer than three applications of a fixed
%	maximally entangling two-qubit gate
%	~\cite{vidal2004universal,vatan2004optimal}.
%	By comparison, an XY- or iSWAP-type interaction has a vanishing
%	third canonical coordinate and can be implemented with two such
%	gates.
%	Independent transpilation of Eq.~(\ref{eq:collision}) into both CZ
%	and CNOT bases, using the compiler's highest optimization level,
%	returned three two-qubit gates in each case.
%	Thus, the three-gate lower bound is attained for the partial-swap
%	unitary considered here.
%	Because this count assumes all-to-all connectivity, it provides a
%	lower bound on the two-qubit gate count for a device that requires
%	routing.
%	The exception is the set of rows with
%	$\gamma_{\rm in}=\pi$, discussed below, for which the
%	within-register operation is equivalent to the identity operator up to a
%	global phase.
%	The quantity \texttt{n2q\_all\_to\_all} is used in
%	Figure~\ref{fig:cost}.

	\item \texttt{n2q\_all\_to\_all} gives the corresponding two-qubit
	gate count in a CNOT/CZ entangling basis for an idealized device with
	all-to-all connectivity.
	For a nontrivial partial swap, this count is
	$3\times\texttt{pswap}$.
	The partial-swap unitary has three nonzero canonical (Weyl)
	coordinates unless $\gamma$ is a multiple of $\pi$, and its standard
	CNOT decomposition therefore requires three entangling
	gates~\cite{vidal2004universal,vatan2004optimal}.
	Because CZ and CNOT are locally equivalent, the same entangling-gate
	count applies to a CZ decomposition.
	Independent transpilation of Eq.~(\ref{eq:collision}) into both CNOT
	and CZ bases, using the compiler's highest optimization level, returned
	three two-qubit gates in each case.
	Thus, \texttt{n2q\_all\_to\_all} should be interpreted as an
	all-to-all CNOT/CZ-equivalent circuit cost rather than as a
	hardware-independent native-gate count.
	The exception is the set of rows with $\gamma_{\rm in}=\pi$,
	discussed below, for which the within-register operation is equivalent
	to the identity up to a global phase.
	The quantity \texttt{n2q\_all\_to\_all} is used in
	Figure~\ref{fig:cost}.
	
	\item \texttt{n2q\_expected} and
	\texttt{n2q\_expected\_hi} incorporate the additional two-qubit
	gates required for routing on a device with restricted
	connectivity.
	They are obtained by applying, respectively, the median and
	90th-percentile routing overheads measured in
	Sec.~\ref{sec:l2methods}.
	Unlike \texttt{pswap} and \texttt{n2q\_all\_to\_all}, these
	routing-inclusive quantities are estimates.	
\end{enumerate}

A fourth circuit-cost quantity is reported only for the measured arms
described in Sec.~\ref{sec:l3}.
For each arm, the state-preparation circuit was transpiled to the native
gate set of IQM Emerald using the selected physical-qubit layout.
The resulting native two-qubit gate count therefore includes routing
overhead and is an exact count for that transpiled circuit, rather than
an estimate.
This quantity is available only for the 10 measured arms.

A separate caveat applies to 544 of the 6,231 L1 cells because the
frontier combines the results of two independent searches.
One search explored depths up to 200 sub-rounds, whereas the other was
limited to $m=0,\ldots,49$ sub-rounds.
For each cell, the selection procedure described in
Sec.~\ref{sec:l1methods} retains the higher-fidelity result.
The file \texttt{frontier\_source.csv} records which search supplied
each selected cell.
The deeper search supplies 5,687 cells, and the shallower search
supplies the remaining 544.

The distinction affects the circuit-cost calculation because the
interaction count depends on whether $\gamma_{\rm in}=0$.
When $\gamma_{\rm in}=0$, the within-register collisions are absent
and the circuit contains fewer partial-swap interactions.
For the 5,687 cells supplied by the deeper search, the deposited
per-candidate data include the corresponding value of
$\gamma_{\rm in}$, so the interaction count can be determined exactly.
For the 544 cells supplied by the shallower search, the deposited data
do not contain the per-candidate collision strengths required to
determine whether $\gamma_{\rm in}=0$.
Their circuit costs are therefore calculated under the conservative
assumption that the within-register collisions are present.
The reported costs for these 544 cells are consequently upper bounds
rather than exact counts.
They can be identified from their source labels in
\texttt{frontier\_source.csv}.

A second caveat concerns 1,003 of the 6,231 cost rows.
According to Eq.~(\ref{eq:collision}), a collision with
$\gamma=\pi$ is equivalent to the identity operator up to a global phase.
Consequently, $\gamma_{\rm in}=\pi$ produces the same state as
$\gamma_{\rm in}=0$, although the former is represented in the circuit
with within-register collisions present.
The search returned $\gamma_{\rm in}=\pi$ for 11 target states:
$D_{7}^{3}$, $D_{8}^{4}$, $D_{9}^{3}$, $D_{10}^{5}$,
$D_{11}^{3}$, $D_{12}^{4}$, $D_{12}^{5}$, $D_{13}^{3}$,
$D_{13}^{4}$, $D_{14}^{3}$, and $D_{14}^{4}$.
Because the \texttt{intra\_on} field distinguishes only between
$\gamma_{\rm in}=0$ and nonzero values, these rows are assigned the
cost of circuits that include the within-register collisions.

As a result, 134,958 of the 11,381,241 two-qubit gates reported in
\texttt{n2q\_all\_to\_all}, corresponding to 1.19\% of the total,
arise from operations that are equivalent to the identity and could
therefore be removed during compilation.
This affects the reported circuit cost but not the prepared state or
its fidelity.
Rows of this type can be identified by testing
\texttt{gamma\_in} against $\pi$.
The script \texttt{code/pi\_l2\_gatecounts.py} performs this check and
reports both the number of affected rows and their associated gate
counts.

The file \texttt{cost\_fidelity\_ladder.json} defines five operating
regimes drawn from the L2 table.
These regimes are summarized in Table~\ref{tab:regimes} and plotted in
Figure~\ref{fig:cost}.
They consist of the best-found operating point from each of the two
searches, a fixed reference parameter set with stopping depths
$m\leq29$, and one full round evaluated with and without
within-register collisions.

The fixed reference set was used to design the hardware campaign
described in Sec.~\ref{sec:l3}.
It contains one set of collision strengths and one stopping depth for
each target state and is provided in
\texttt{published\_parameters.md}.
Entries labeled \texttt{published} in the data tables refer to this
reference set.
The reference points should not be interpreted as the best-found
points in the deeper search.
For example, for $D_{5}^{2}$ the reference depth is $m=26$, whereas
the deeper search reaches its best-found fidelity at $m=174$.

Across the 31 target states, the median value of
$F_{\rm one\mbox{-}round}-F_{\rm deep}$ is $-0.0257$, with the largest
decrease equal to $-0.1349$.
At the same time, the median ratio of the two-qubit gate count of the
deep-search operating point to that of one full round with
$\gamma_{\rm in}=0$ is 72.9, with a maximum ratio of 173.
Figure~\ref{fig:cost} summarizes this fidelity--cost trade-off for all
five operating regimes and all 31 target states.

\begin{table}[tb]
\caption{\label{tab:regimes}
	Five operating regimes defined in
	\texttt{cost\_fidelity\_ladder.json}, summarized by their medians
	across the 31 target states.
	Here $m$ is the depth in sub-rounds, pSWAP is the reported
	partial-swap interaction cost, and 2q is the corresponding all-to-all
	CNOT/CZ-equivalent two-qubit gate cost, $3\times\mathrm{pSWAP}$.}
  \begin{ruledtabular}
  \footnotesize
  \begin{tabular}{@{}l r r r r r@{}}
    operating point & states & $F$ & $m$ & pSWAP & 2q \\
    \colrule
    deep-search best found & 31 & 0.8869 & 116 & 944 & 2832 \\
    best found within $m \leq 49$ & 31 & 0.8633 & 4 & 78 & 234 \\
    cap-30 reference set, $m \leq 29$ & 31 & 0.8559 & 4 & 64 & 192 \\
    one full round, $\gamma_{\rm in}$ free & 31 & 0.8504 & 3 & 55 & 165 \\
    one full round, $\gamma_{\rm in} = 0$ & 31 & 0.8446 & 3 & 22 & 66 \\
  \end{tabular}
  \end{ruledtabular}
\end{table}

\subsection{L3: hardware measurements}
\label{sec:l3}

The hardware measurements were performed on IQM Emerald, a 54-qubit
superconducting processor with 85 calibrated two-qubit couplings,
accessed through Amazon Braket in the \texttt{eu-north-1} region.
The calibration snapshot recorded at the time of submission,
\texttt{2026-08-20T05:19:15.989291+00:00}, is included in the
released data.
In this snapshot, the calibrated two-qubit gate error ranges from
0.153\% to 8.206\%, corresponding to a factor of 54 across the
device, with a median of 0.349\%.
The median relaxation and dephasing times are
$T_1=54.7~\mu$s and $T_2=16.6~\mu$s, respectively.

\begin{figure*}[t!]
  \includegraphics[width=1.5\columnwidth]{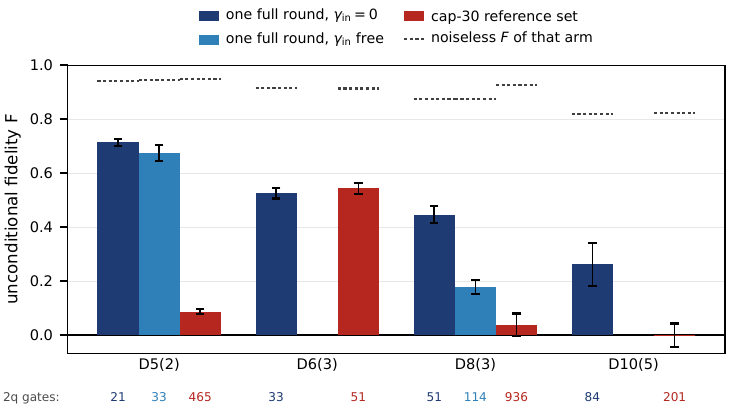}
%  \caption{\label{fig:arms}
%  	Estimated unconditional fidelity $F$ for each measured hardware arm.
%  	The horizontal dash above each bar gives the noiseless fidelity
%  	$F_{\rm ideal}$ of the corresponding circuit, and the native
%  	two-qubit gate count is shown below each bar.
%  	Error bars show the quoted uncertainty $\sigma_F$ obtained from the
%  	error-bar analysis described in Sec.~\ref{sec:fidelity}.}
	\caption{\label{fig:arms}
		Estimated unconditional fidelity $F$ for each measured hardware arm.
		The horizontal dash above each bar indicates the noiseless fidelity
		$F_{\rm ideal}$ of the corresponding circuit, and the native
		two-qubit gate count is shown beneath each bar.
		Error bars show the quoted uncertainty $\sigma_F$ obtained from the
		error-bar analysis described in Sec.~\ref{sec:fidelity}.
	}
\end{figure*}

The file \texttt{arm\_results.csv} contains results for 10 circuit
arms across four target states.
The arms are listed in Table~\ref{tab:arms} and shown in
Figure~\ref{fig:arms}.
Here, an arm denotes one choice of circuit parameters for a given
target state.
The \texttt{gin\_zero} arm uses one full round with
$\gamma_{\rm in}=0$, whereas \texttt{gin\_free} uses one full round
with $\gamma_{\rm in}$ optimized freely.
The \texttt{published} arm corresponds to the cap-30 reference set
defined in Sec.~\ref{sec:l2}.

The three arm types were not implemented for every target state.
For $D_{5}^{2}$ and $D_{8}^{3}$, all three arms were measured.
For $D_{6}^{3}$ and $D_{10}^{5}$, only two arms were measured because
the reference set already terminates after one full round with
$\gamma_{\rm in}$ free.
For these two states, an additional \texttt{gin\_free} arm would
therefore have differed from the \texttt{published} arm only in the
collision strengths.
Comparisons involving all three arm types are consequently restricted
to $D_{5}^{2}$ and $D_{8}^{3}$.

For each target state, all measured arms were assigned to the same
set of physical qubits within a single hardware session.
This design removes differences in qubit layout as a source of
variation when comparing arms of the same state.
The session consisted of 440 circuits grouped into 20 tasks and
included 88,656 shots in total.
The tasks were submitted at
\texttt{2026-08-21T00:34:00.279745+00:00} using Amazon Braket
verbatim mode, at a total execution cost of \$167.35.

Figure~\ref{fig:arms} compares the measured fidelity $F$ with the
noiseless fidelity $F_{\rm ideal}$ of the same circuit.
The ratio $F/F_{\rm ideal}$ therefore quantifies the fraction of the
ideal fidelity retained on hardware.
For the two states with all three arm types, the deeper reference arms
retain substantially smaller fractions of their noiseless fidelities
than the one-round arms.
For $D_{5}^{2}$, the 26-sub-round reference arm gives
$F/F_{\rm ideal}=0.091$, whereas the two one-round arms give values
between 0.713 and 0.758.
For $D_{8}^{3}$, the corresponding values are 0.041 for the reference
arm and 0.204--0.510 for the one-round arms.

Each arm reports the fidelity $F=\langle D_N^K|\rho|D_N^K\rangle$ 
of the prepared state $\rho$ to the ideal state $|D_N^K\rangle$, together with its quoted uncertainty $\sigma_F$.
We also report the population of the target excitation sector,
$P_K=\mathrm{Tr}(\Pi_K\rho)$, where $\Pi_K$ is the projector onto the
subspace containing exactly $K$ excitations.
The ratio $F/P_K$ and the noiseless fidelity $F_{\rm ideal}$ of the
corresponding circuit are reported as well.
The latter is used to determine the fraction of the noiseless fidelity
retained on hardware.

The ratio $F/P_K$ has a direct interpretation as an ideal conditional
fidelity.
Since the target state lies entirely in the $K$-excitation subspace,
$\Pi_K|D_N^K\rangle=|D_N^K\rangle$.
If an ideal projective measurement of the total excitation number
selects the outcome $K$, the post-selected state has fidelity
\begin{equation}
	\langle D_N^K |
	\frac{\Pi_K \rho \Pi_K}{\mathrm{Tr}(\Pi_K \rho)}
	| D_N^K \rangle
	=
	\frac{\langle D_N^K|\rho|D_N^K\rangle}{P_K}
	=
	\frac{F}{P_K}.
	\label{eq:heralded}
\end{equation}
Equation~(\ref{eq:heralded}) therefore gives the fidelity conditioned
on an ideal, nondestructive measurement of the excitation number.
Such a heralding measurement was not implemented in the present
experiment.
In particular, determining the excitation number by measuring all
qubits in the computational basis would destroy the coherences that
distinguish $|D_N^K\rangle$ from an incoherent mixture of basis states
with the same excitation number.
Moreover, a physical heralding procedure would introduce additional
measurement errors.
Thus, $F/P_K$ should be interpreted as an idealized conditional
fidelity rather than as the result of a directly implemented heralding
protocol.
Because the projector onto the target state is contained within the
$K$-excitation subspace, $F\leq P_K$; this inequality is satisfied by
all 10 measured arms.

Table~\ref{tab:arms} contains one negative estimate of $F$.
The physical fidelity
$\langle D_N^K|\rho|D_N^K\rangle$ is non-negative, but its estimator is
not constrained to the interval $[0,1]$.
%For each arm, four independent fidelity estimates are obtained from
%four disjoint replicates of the measurement settings.
%The reported value of $F$ is their mean, and the quoted uncertainty is
%the standard error of that mean.
%For $D_{10}^{5}$/\texttt{published}, the four replicate estimates are
%$+0.0569$, $-0.0822$, $+0.0322$, and $-0.0276$.
%These replicate values illustrate the statistical fluctuations of the
%fidelity estimator; two of the four replicates yield negative values.
%The published central estimate $F=-0.0014$ is obtained from the primary
%estimator run and is not the arithmetic mean of the four replicate
%values listed above.
As part of the uncertainty analysis, four replicate fidelity estimates
are obtained from disjoint subsets of the measurement settings.
These replicate values characterize the statistical fluctuations of the
estimator but do not define the published central fidelity $F$, which is
obtained from the primary estimator run described in
Sec.~\ref{sec:fidelity}.
For $D_{10}^{5}$/\texttt{published}, the four replicate estimates are
$+0.0569$, $-0.0822$, $+0.0322$, and $-0.0276$.
Two of the four replicate estimates are negative, illustrating that the
unconstrained estimator can fluctuate below zero when the fidelity is
close to the statistical noise floor.

The fidelity estimator is a signed weighted sum of measured Pauli
expectation values.
The signs of the weights are determined by the corresponding
expectation values of the target Dicke state, so statistical
fluctuations in the measured terms can produce both positive and
negative contributions.
Although the estimator is unbiased, it is not explicitly constrained
to the physical interval $[0,1]$.
This effect becomes important when the true fidelity is close to zero.
The linear inversion used for per-qubit readout-error correction can
further increase the statistical spread.

For the $D_{10}^{5}$/\texttt{published} arm, the estimated fidelity is
$F=-0.0014\pm0.0437$, placing zero only approximately $0.03$ standard
errors from the estimated mean.
Its measured excitation-sector population is $P_K=0.2271$.
For comparison, a maximally mixed ten-qubit state has population $\binom{10}{5} / 2^{10}=0.2461$
in the five-excitation subspace.
These observations are consistent with a state whose fidelity to the
target has fallen to the statistical noise floor.
We therefore report the fidelity estimate without clipping it to zero,
since imposing non-negativity on the estimator would introduce an
upward bias for fidelities near zero.

\begin{table*}[tb]
\caption{\label{tab:arms}
	Measured hardware arms.
	$n_{2q}$ is the native two-qubit gate count after transpilation,
	$F_{\rm ideal}$ is the noiseless fidelity of the same circuit, and
	$F$ is the estimated unconditional fidelity with its quoted
	uncertainty $\sigma_F$ from the error-bar analysis.
	The ratio $F/F_{\rm ideal}$ gives the fraction of the noiseless
	fidelity retained on hardware.
	$P_K$ is the population of the target $K$-excitation sector, and
	$F/P_K$ is the ideal conditional fidelity defined by
	Eq.~(\ref{eq:heralded}).
	The column ``rounds'' equals the number of sub-rounds divided by $K$.
	The physical qubits used for each arm are listed in
	\texttt{arm\_results.csv}.}
  \begin{ruledtabular}
  \footnotesize
  \begin{tabular}{lccccccccc}
	state & arm & sub-rounds & rounds & $n_{2q}$ & $F_{\rm ideal}$ &
	$F$ & $F/F_{\rm ideal}$ & $P_K$ & $F/P_K$ \\
	\colrule
	$D_{5}^{2}$ & \texttt{gin\_zero} & 2 & 1 & 21 & 0.9413 & $+0.7138 \pm 0.0119$ & 0.758 & 0.8247 & +0.8655 \\
	$D_{5}^{2}$ & \texttt{gin\_free} & 2 & 1 & 33 & 0.9465 & $+0.6746 \pm 0.0285$ & 0.713 & 0.7753 & +0.8701 \\
	$D_{5}^{2}$ & \texttt{published} & 26 & 13 & 465 & 0.9494 & $+0.0865 \pm 0.0079$ & 0.091 & 0.3634 & +0.2380 \\
	$D_{6}^{3}$ & \texttt{gin\_zero} & 3 & 1 & 33 & 0.9150 & $+0.5248 \pm 0.0192$ & 0.574 & 0.6679 & +0.7857 \\
	$D_{6}^{3}$ & \texttt{published} & 3 & 1 & 51 & 0.9137 & $+0.5439 \pm 0.0203$ & 0.595 & 0.6721 & +0.8093 \\
	$D_{8}^{3}$ & \texttt{gin\_zero} & 3 & 1 & 51 & 0.8751 & $+0.4463 \pm 0.0313$ & 0.510 & 0.6524 & +0.6841 \\
	$D_{8}^{3}$ & \texttt{gin\_free} & 3 & 1 & 114 & 0.8760 & $+0.1785 \pm 0.0259$ & 0.204 & 0.4443 & +0.4017 \\
	$D_{8}^{3}$ & \texttt{published} & 24 & 8 & 936 & 0.9266 & $+0.0380 \pm 0.0414$ & 0.041 & 0.2622 & +0.1449 \\
	$D_{10}^{5}$ & \texttt{gin\_zero} & 5 & 1 & 84 & 0.8184 & $+0.2612 \pm 0.0801$ & 0.319 & 0.5012 & +0.5211 \\
	$D_{10}^{5}$ & \texttt{published} & 5 & 1 & 201 & 0.8226 & $-0.0014 \pm 0.0437$ & -0.002 & 0.2271 & -0.0061 \\
\end{tabular}
  \end{ruledtabular}
\end{table*}

Table~\ref{tab:arms} reports the hardware measurements used in this
study.
An earlier superseded session is retained in
\texttt{data/raw/} under the prefix \texttt{first\_run\_}$\ast$ for
provenance and verification purposes.
The file \texttt{cross\_session\_check.json} contains the hash
comparison showing that the circuits executed in the two sessions are
not identical.
In addition, the two sessions used different assignments of physical
qubits.
The earlier session is therefore not suitable for arm-by-arm
comparison with the measurements in Table~\ref{tab:arms} and is not
included in the reported hardware results.

\section{Methods}
\label{sec:methods}

\subsection{Frontier Construction}
\label{sec:l1methods}

Each frontier cell is obtained through two distinct optimization
stages.
The first stage optimizes the collision strengths
$\gamma_{\rm in}$ and $\gamma_{\rm sh}$, whereas the second optimizes
the single-qubit phase layer.
Because the two stages involve different parameters and optimization
methods, they are described separately below.

\emph{The collision strengths.}
The collision strengths $\gamma_{\rm in}$ and $\gamma_{\rm sh}$ are
optimized using the multi-start procedure introduced in
Ref.~\cite{vu2026intelligent}.
For each target state, 256 initial parameter pairs are chosen on a
uniform $16\times16$ grid with spacing 0.2~rad over
$\gamma_{\rm in}\in[0,\pi]$ and
$\gamma_{\rm sh}\in[0.01,\pi]$.
Each grid point initializes an independent L-BFGS-B
optimization~\cite{byrd1995limited}, using finite-difference gradients
and the same parameter bounds.
For every target state, the initial parameter pair, the refined pair,
and the final loss for all 256 candidates are provided in
\texttt{cap200\_starts/}.

\emph{The phase layer.}
The single-qubit phase-rotation angles are optimized separately for
each candidate and at each circuit depth.
This higher-dimensional optimization is performed using batched
Adam~\cite{kingma2014adam}, with four independent restarts of
600 steps each.
The results were cross-checked against an independent GPU
implementation.

The final operating point for each state is selected according to the
post-phase fidelity, with the comparison performed over both candidate
index and circuit depth.
The order of these operations matters.
If the circuit depth is fixed before optimizing the phase layer, the
selected fidelity changes for 8 of the 31 states, with a maximum
difference of $0.068$.
Under the selection rule used here, the median depth of the
best-found operating point across the 31 states is 116 sub-rounds.

Two independent searches were performed over the same collision-strength
parameter space.
One search explored depths up to 200 sub-rounds, whereas the other covered $m=0,\ldots,49$ sub-rounds.
Because the searches were performed independently, the deeper search
does not necessarily reproduce or improve upon every result of the
shallower search.
Among the 31 per-state best-found fidelities, 30 are supplied by the
deeper search and one by the shallower search.

The two search outputs are combined in two stages.
First, the results are compared cell by cell.
The shallower-search value replaces the deeper-search value only when
its fidelity is larger by more than $10^{-5}$.
Smaller differences are treated as numerically indistinguishable,
because the searches are independent and the stored arrays use
32-bit floating-point precision.

Second, a cumulative maximum is taken over circuit depth.
As defined in Sec.~\ref{sec:depth}, a budget row includes all circuits
with depths up to its corresponding maximum depth.
The resulting fidelity frontier must therefore be nondecreasing with
increasing budget.
If a frontier cell inherits its value from a smaller depth, its
provenance is recorded as
\texttt{\textless{}search\textgreater{}@B\textless{}=n}, indicating
that the selected value was obtained at an earlier budget rather than
at the current depth.
Both original search outputs are included in the deposit, allowing the
combined frontier to be reconstructed using a different selection rule.

\subsection{Derivation of the circuit cost}
\label{sec:l2methods}

For rows with complete collision-strength information, the number of
partial-swap interactions is determined analytically from $N$, $K$,
the number of sub-rounds, and whether the within-register collision
strength is nonzero.
The resulting closed-form expression reproduces every stored
transpiled circuit exactly.
For the 544 cells supplied by the shallower search, the value of
$\gamma_{\rm in}$ required to determine the exact interaction count is
not available; their reported interaction counts are therefore
conservative upper bounds, as described in Sec.~\ref{sec:l2}.

As defined in Sec.~\ref{sec:l2},
\texttt{n2q\_all\_to\_all} assigns three CNOT/CZ-equivalent
two-qubit gates to each nontrivial partial-swap interaction.
A device with restricted connectivity can require additional
two-qubit gates for routing.
We estimated this routing overhead from 150 transpiled circuits.
Relative to the all-to-all gate count, the measured overhead has a
median of 25.0\% and a 90th percentile of 56.7\%.
These two values are used to construct the routing-inclusive cost
estimates reported for each frontier cell.

A device with restricted qubit connectivity requires additional
two-qubit gates for routing.
We estimated this routing overhead from 150 transpiled circuits.
Relative to the all-to-all gate count, the measured overhead has a
median of 25.0\% and a 90th percentile of 56.7\%.
These two values are used to construct the routing-inclusive cost
estimates reported for each frontier cell.

\subsection{Fidelity estimation}
\label{sec:fidelity}

Measurements in the computational basis alone are insufficient to
determine the fidelity to a Dicke state.
The ideal Dicke state and an incoherent mixture of the same
$K$-excitation basis states have identical computational-basis
probability distributions and the same excitation-sector population
$P_K$.
They differ, however, in their off-diagonal coherences and therefore
in their fidelities to the target state.
For example, the uniform incoherent mixture over the
$\binom{5}{2}=10$ basis states in the two-excitation subspace has
fidelity $1/\binom{5}{2}=0.1$ with $|D_5^2\rangle$, whereas the ideal
Dicke state has unit fidelity.
Accessing the required coherences therefore requires measurements in
bases other than the computational basis.
For this reason, each state-preparation circuit was followed by
multiple measurement-basis rotations involving the $X$, $Y$, and $Z$
bases.

Rather than reconstructing the full density matrix, the fidelity
estimator exploits the permutation symmetry of the Dicke target
state~\cite{toth2010permutationally}.
Pauli terms related by qubit permutations are grouped into symmetry
orbits, following the general strategy of direct fidelity
estimation~\cite{flammia2011direct,da2011practical}.
This reduces the number of required measurement settings relative to
full tomography.
Table~\ref{tab:orbits} compares the corresponding circuit counts with
those required for full state tomography.
Readout errors are corrected independently for each qubit using the
assignment-error matrices contained in the calibration snapshot.

For each arm, the uncertainty analysis uses four independent
measurement-setting replicates.
Because the fidelity estimator samples the Pauli terms used in the
reconstruction, the error-bar analysis additionally repeats the
estimator over 32 independent support draws.
The central fidelity $F$ reported in \texttt{arm\_results.csv} is the
result of the primary estimator run and is therefore not defined as the
arithmetic mean of the four values listed in
\texttt{replicates.csv}.
The quoted uncertainty $\sigma_F$ is obtained from the corresponding
error-bar analysis.
No parametric noise model is assumed in this procedure.
The use of four measurement-setting replicates implies three degrees of
freedom for statistics based on the replicate scatter; this limitation
is discussed further in Sec.~\ref{sec:limitations}.

\begin{table}[tb]
	\caption{\label{tab:orbits}
		Number of measurement circuits required per hardware arm.
		Full tomography requires $3^N$ measurement settings.
		The column ``orbits'' gives the number of Pauli terms after grouping
		under permutations of the target state, ``settings'' gives the number
		of distinct measurement settings per replicate, and ``circuits''
		gives the total number used in the four-replicate uncertainty
		analysis.}
  \begin{ruledtabular}
  \begin{tabular}{lrrrr}
    state & full tomography & orbits & settings & circuits \\
    \colrule
    $D_{5}^{2}$ & 243 & 20 & 6 & 24 \\
    $D_{6}^{3}$ & 729 & 20 & 10 & 40 \\
    $D_{8}^{3}$ & 6,561 & 46 & 10 & 40 \\
    $D_{10}^{5}$ & 59,049 & 56 & 21 & 84 \\
  \end{tabular}
  \end{ruledtabular}
\end{table}

\subsection{Experimental design}
\label{sec:design}

The calibrated two-qubit gate error of IQM Emerald varies by a factor
of 54 across device couplings.
The choice of physical qubits was therefore treated as part of the
experimental design.
A 14-qubit working region was selected by minimizing the summed
two-qubit gate error after excluding three qubits from consideration.
Within this region, the physical-qubit layout for each target state
was chosen by minimizing the summed gate error rather than the
transpiled gate count.
A controlled layout test was performed twice for the same circuit arm.
Changing the physical-qubit layout changed the conditional fidelity
$F/P_K$ by 0.148 and 0.100 in the two runs
(\texttt{layout\_experiment.json}).
These differences are larger than most of the fidelity differences
observed between circuit arms.
To control for this layout dependence, all arms corresponding to a
given target state were executed on the same set of physical qubits.

\begin{figure*}[t!]
  \includegraphics[width=2\columnwidth]{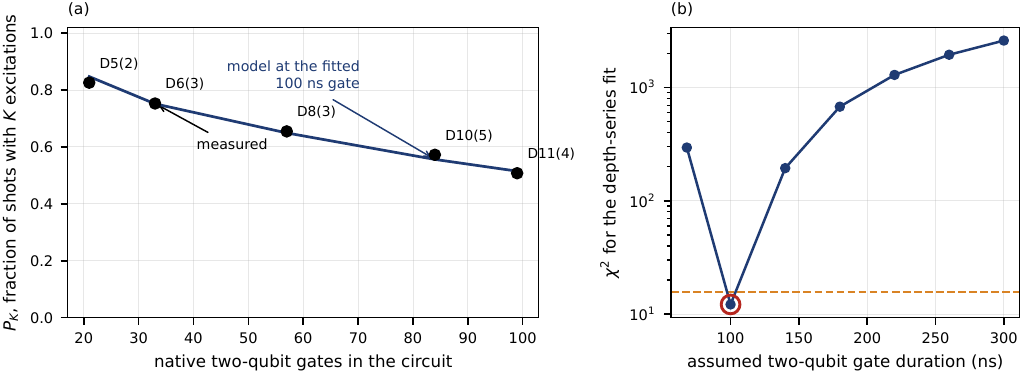}
%	\caption{\label{fig:pk}
%		Effective two-qubit gate-duration fit.
%		(a) Measured excitation-sector population $P_K$ versus native
%		two-qubit gate count, together with the model evaluated at the fitted
%		effective gate duration of 100~ns.
%		(b) $\chi^2$ of the depth-series model as the assumed two-qubit gate
%		duration is varied.
%		The dashed horizontal line denotes the alternative fit obtained by
%		rescaling the calibrated two-qubit gate errors at a fixed duration of
%		68~ns.}
	\caption{\label{fig:pk}
		Effective two-qubit gate-duration fit.
		(a) Measured excitation-sector population $P_K$ versus the native
		two-qubit gate count for the preliminary depth series.
		The solid curve shows the model evaluated at the fitted effective
		two-qubit gate duration of 100~ns.
		(b) Gate-duration scan obtained by evaluating $\chi^2$ for the same
		depth-series measurements as the assumed two-qubit gate duration is
		varied.
		The minimum occurs at 100~ns, with $\chi^2=12.2$.
		The dashed horizontal line shows the alternative fit obtained by
		fixing the duration at 68~ns and rescaling all calibrated two-qubit
		gate errors by a common factor.
		The best such fit requires a factor $\lambda=2.4$ and gives
		$\chi^2=15.7$.
	}
\end{figure*}

Circuit duration was characterized before selecting states for the
tomography measurements because decoherence accumulates with circuit
runtime.
The calibration snapshot provides gate fidelities, $T_1$, $T_2$, and
readout errors, but does not specify gate durations.
We therefore estimated the effective two-qubit gate duration directly
from hardware measurements.
Because the excitation-sector population $P_K$ can be obtained from a
single all-$Z$ measurement setting, it was used for this preliminary
depth scan.
For each state, measurements were performed at five circuit depths
spanning 21 to 99 native two-qubit gates, with 4000 shots per task and
a total execution cost of \$37.00.
The assumed two-qubit gate duration was varied from 68 to 300~ns while
the calibrated gate fidelities were held fixed.
The minimum $\chi^2$ occurs at 100~ns, with
$\chi^2=12.2$ for four degrees of freedom.
For comparison, $\chi^2=295.5$ at 68~ns and
$\chi^2=194.5$ at 140~ns.
Figure~\ref{fig:pk} shows both the measured depth dependence and the
duration scan.

The measured decrease of $P_K$ with circuit depth is steeper than
predicted using the calibration data alone.
We considered two possible explanations for this discrepancy.
The first is a longer effective two-qubit gate duration, which
increases decoherence over the circuit runtime.
The second is a larger two-qubit gate error than reported in the
calibration snapshot.
To test the latter possibility, the gate duration was fixed at
68~ns and every calibrated two-qubit gate error was multiplied by a
common factor $\lambda$.
The best fit gives $\lambda=2.4$, corresponding within this model to
two-qubit errors 2.4 times larger than the calibrated values.
This model gives $\chi^2=15.7$, compared with
$\chi^2=12.2$ obtained by fitting the gate duration alone.
All hardware-performance predictions reported below use the fitted
effective gate duration.

The fitted model was then used to select target states for the
tomography measurements.
States were selected when the lower end of the predicted
post-selected-fidelity range exceeded 0.5.
For each state, this range spans the predictions obtained from the two
model variants described above.
For the largest measured state, $D_{10}^{5}$, the predicted
post-selected fidelity lies in the interval $[0.532,0.542]$.
By comparison, $D_{11}^{4}$ has a predicted interval of
$[0.491,0.503]$ and was not selected for measurement.
The model was used only for experimental design and circuit selection;
it does not enter the estimation of any experimental observable
reported in this work.

\subsection{Consistency of the released tables}
\label{sec:validation}

We performed several independent validation checks on the released
quantities.
Released frontier cells were reconstructed from the recorded collision
strengths using an independent Qiskit implementation, and the phase
layer was re-optimized with SciPy. This validation path is independent
of the GPU simulator and batched optimizer used to generate the
released frontier. The largest disagreement is $4\times10^{-6}$, which is
the storage precision of the 32-bit arrays. Reading the published
table back through a second independent Qiskit path agrees to
$5\times10^{-7}$.

For rows with complete collision-strength information, the closed-form
interaction count reproduces every stored transpiled circuit exactly.
Independent transpilation into both CNOT and CZ bases returns three
entangling gates for each nontrivial partial swap, consistent with the
all-to-all cost definition of Sec.~\ref{sec:l2}.
The 544 cells supplied by the shallower search are excluded from this
exact-count statement because their reported costs are conservative
upper bounds.

$P_K$ was recomputed from the raw counts for every arm of both
sessions on a separate implementation and agrees after rounding to four decimal places. The two sessions' circuits hash differently,
which is the basis for the instruction in Sec.~\ref{sec:usage} not
to pool them.

The arrays also satisfy the following invariant, which constrains their interpretation. Before the phase layer, the overlap with the target
equals the value it would take for an equal-weight mixture over the
sector,
\begin{equation}
  |\langle D_N^K | \psi_{\rm pre} \rangle|^2
  = \binom{N}{K}^{-1},
  \label{eq:invariance}
\end{equation}
at all 1,595,136 stored candidate-depth pairs, with a largest
relative deviation of $6\times10^{-8}$. This invariant provides an independent consistency check on the
pre-phase states across all stored candidate--depth pairs.

The deposit includes the scripts for these and the remaining internal checks under \texttt{code/}, each paired with a test that breaks the property it checks and confirms that the check fails.

\section{Limitations}
\label{sec:limitations}

Each item below states what the data do not support, and what a user
should do about it.

\begin{enumerate}
\item \emph{L1 is noiseless, so it does not predict hardware.} An
  L1 value is what the circuit would reach on a perfect device. 
    On Emerald the best of the 10 measured arms retained
  0.758 of its noiseless fidelity and the deepest approached the value returned by a fully decohered circuit 
  (Table~\ref{tab:arms}, $F/F_{\rm ideal}$).
   Therefore, L1 fidelities should not be interpreted as expected from an experimental measurement.
	\item \emph{L1 reports best-found search values, not proven optima.}
	Each frontier cell is produced by the released searches and the
	selection rule of Sec.~\ref{sec:l1methods}.
	It therefore gives a lower bound on the best fidelity that the protocol
	may achieve within that depth budget, rather than a certified optimum.
	A better search can raise the frontier, so small differences between
	cells may reflect the search procedure rather than the protocol itself.
\item \emph{544 of the 6,231 cost rows are upper bounds,
	not exact counts.}
These 544 rows correspond to frontier cells selected from the shallower
search. For these cells, the value of the within-register collision
strength $\gamma_{\mathrm{in}}$ is not available in the released
per-candidate data. Their circuit costs are therefore calculated under
the conservative assumption that within-register collisions are present
(Sec.~\ref{sec:l2}).
The 544 affected rows are identified by their shallower-search
provenance in \texttt{frontier\_source.csv}.
The exact costs of these 544 rows cannot be reconstructed from
the deposit because the approximately 1.4~GB per-candidate output of
the shallower search is not included.
\item \emph{The routing columns are extrapolated at large depth.}
  The overhead was fitted on 150 circuits far
  shallower than the deep end of the frontier. At depths of order
  200 the routing-inclusive columns carry no measured support.
  \texttt{pswap} and \texttt{n2q\_all\_to\_all} are unaffected.
\item \emph{The hardware layer is one session on one processor.}
  Four target states and 10 arms were measured.
  The uncertainty analysis uses four measurement-setting replicates;
  statistics based on their replicate scatter therefore have three
  degrees of freedom. No arm was repeated with an identical circuit in a second
  session, so run-to-run reproducibility of a single arm is not
  established. The numbers characterize the circuits on a specific
  qubit set on a specific day. In general, they characterize neither the device nor the protocol.
\item \emph{$F$ and $F/P_K$ quantify different preparation conditions.}
	The quantity $F$ is the unconditional fidelity of the prepared state and
	therefore refers to all experimental shots.
	By Eq.~(\ref{eq:heralded}), $F/P_K$ is the fidelity conditioned on the
	system being found in the target $K$-excitation subspace, whose measured
	population is $P_K$.
	This conditional quantity corresponds to the fidelity that would be obtained
	after an ideal excitation-number herald.
	Such a heralding measurement was not implemented in the present experiment;
	therefore, $F/P_K$ is inferred from the measured $F$ and $P_K$ rather than
	obtained from a separate heralded measurement.
	A physical heralding procedure would introduce additional experimental errors.
	Consequently, comparisons with other preparation schemes should use $F$ for
	unconditional protocols and $F/P_K$ only for heralded protocols.
\item \emph{Depth stops at 200 sub-rounds.} That is
  $40$ full rounds at $K=5$ and
  $100$ at $K=2$, so the deepest point in the release is
  not a fixed number of full rounds across states.
  The dataset therefore does not support conclusions about behavior beyond 200 sub-rounds.
\end{enumerate}

\section{Usage notes}
\label{sec:usage}

Section~\ref{sec:reuse} sets out what the layers were assembled to
support; this section covers the mechanics. Start from
\texttt{README.md} for the layout and
\texttt{DATA\_DICTIONARY.md} for column definitions. For the frontier,
read \texttt{frontier\_canonical.csv} and check
\texttt{frontier\_source.csv} for provenance before quoting a cell. For
circuit cost, join on (state, \texttt{budget}).

The two layers can be combined to determine the maximum depth and corresponding fidelity achievable within a fixed two-qubit gate budget:

\begin{small}
	\begin{verbatim}
		import pandas as pd
		
		L1 = pd.read_csv("data/L1_frontier/"
		"frontier_canonical.csv")
		L2 = pd.read_csv("data/L2_gate_costs/"
		"gatecounts_frontier.csv")
		
		state, budget_2q = "D8_3", 500
		
		# rows for this state that fit 
		#   the reported gate budget
		c = L2[L2.state == state]
		c = c[c.n2q_all_to_all <= budget_2q]
		
		B = int(c.budget.max())   # deepest that fits
		row = c[c.budget == B].iloc[0]
		
		print(B - 1,         # depth m, sub-rounds
		row.pswap,           # partial swaps
		row.n2q_all_to_all,  # all-to-all 2q cost
		L1.loc[B - 1, state])# fidelity there
	\end{verbatim}
\end{small}

\noindent
The subtraction is the 1-based budget column of
Sec.~\ref{sec:depth}: row $B$ covers depths $m \le B-1$. For 544 frontier cells supplied by the shallower search, the reported
gate count is a conservative upper bound because the corresponding
collision-strength information is unavailable.
Consequently, a budget calculation involving one of these cells is
conservative rather than exact.
The provenance of each frontier cell is recorded in
\texttt{frontier\_source.csv}.

A small discrepancy can arise when comparing \texttt{replicates.csv}  with the published \(F\) values.
That table's four values per arm come from the error-bar study in
\texttt{error\_bars.json}, which re-runs the estimator with its own
support draw, so averaging them does not return the published $F$ to
the last digit. The estimator samples which Pauli terms to use, so two
runs over the same counts differ slightly; that spread is recorded per
arm as \texttt{F\_seed\_spread}. Across the 10 arms the
difference between the replicate mean and the published $F$ reaches
0.0144, in every case smaller than that arm's own quoted
error bar. Use \texttt{arm\_results.csv} for the value of $F$ and
\texttt{replicates.csv} to see the scatter behind it.

Four further points affect how the tables should be used.
Budget, sub-round, and full round are distinct quantities, as defined in Sec.~\ref{sec:protocol}, and should not be conflated. A fourth count appears if Ref.~\cite{vu2026intelligent} is read
alongside this one: what it calls a round is what this release calls a
sub-round, so a depth quoted from that paper is already in sub-rounds
and must not be multiplied by $K$. 
Prefer the unconditional $F$ to $F/P_K$ unless heralding is part of
the protocol being compared, because the ratio assumes an operation
this experiment did not perform. Do not pool the superseded session with the reported one: the two
sessions used different physical-qubit assignments, and the circuit
hashes differ. Routing-inclusive gate counts should be treated as estimates, whereas partial-swap counts are exact where the required collision-strength information is available.

\texttt{reproduce.py} regenerates every processed file from the raw
counts, and the environment is pinned in \texttt{requirements.txt}:
the analysis chain needs only NumPy, SciPy, Qiskit and Qiskit Aer at
the versions listed there, and Matplotlib is required by the figure
script alone.

\begin{acknowledgments}
F.O. acknowledges financial support from Tokyo International University Personal Research Fund and Special Grant-in-Aid for Research Work. 
During preparation of this manuscript, the authors used OpenAI ChatGPT (GPT-5.6 Sol; accessed August 2026) to assist with manuscript organization, language revision, and drafting of selected explanatory passages. The authors provided the relevant source material and section-specific instructions, reviewed and revised all AI-assisted text, and independently verified the scientific statements, numerical values, references, and conclusions. The authors take full responsibility for the final content.
\end{acknowledgments}

\section*{Data availability}

The data described in this article are openly available in Zenodo at
\url{https://doi.org/10.5281/zenodo.22095760}, reference number
10.5281/zenodo.22095760.
% APS pre-scripted statement (Aug 2024 policy). Replace the
% placeholder DOI with the minted one before submission.

%The data described in this article are openly available in Zenodo at
%\url{https://doi.org/10.5281/zenodo.XXXXXXX}, reference number
%10.5281/zenodo.XXXXXXX. The deposit is a single archive,
%\texttt{<ARCHIVE\_NAME>.zip}, of <SIZE> holding <N> files in the
%directories of Table~\ref{tab:inventory}, with SHA-256 checksum
%
%\begin{small}
%	\begin{verbatim}
%		<64 hexadecimal characters>
%	\end{verbatim}
%\end{small}
%
%\noindent
%and per-file checksums in \texttt{MANIFEST.sha256}. Tabular data are
%CSV with a UTF-8 header row, the optimization surfaces are
%\texttt{.npy} arrays at 32-bit precision, and the instrument records,
%fits and verification outputs are JSON;
%\texttt{DATA\_DICTIONARY.md} defines every column of every table.
%
%The data are released under the Creative Commons Attribution 4.0
%International license (CC BY 4.0). The code is released separately
%under the MIT license, as stated below.

\section*{Code availability}

The analysis and verification chain is included in the same deposit
under \texttt{code/}, released under the MIT license, with the
environment pinned in \texttt{requirements.txt}. The entry points are
\texttt{reproduce.py} to regenerate every processed file from the raw
counts, \texttt{pi\_verify\_frontier.py} for the frontier,
\texttt{pi\_raw\_invariant.py} for the invariance check,
\texttt{pi\_verify\_all.py} for the hardware layer, and
\texttt{pi\_figures.py} to redraw the figures.

%\bibliography{manuscript}
%\bibliography{C:/Users/seval/Dropbox/OQuL/OQuLBib/OQuL}	

\begin{thebibliography}{67}%
	\makeatletter
	\providecommand \@ifxundefined [1]{%
		\@ifx{#1\undefined}
	}%
	\providecommand \@ifnum [1]{%
		\ifnum #1\expandafter \@firstoftwo
		\else \expandafter \@secondoftwo
		\fi
	}%
	\providecommand \@ifx [1]{%
		\ifx #1\expandafter \@firstoftwo
		\else \expandafter \@secondoftwo
		\fi
	}%
	\providecommand \natexlab [1]{#1}%
	\providecommand \enquote  [1]{``#1''}%
	\providecommand \bibnamefont  [1]{#1}%
	\providecommand \bibfnamefont [1]{#1}%
	\providecommand \citenamefont [1]{#1}%
	\providecommand \href@noop [0]{\@secondoftwo}%
	\providecommand \href [0]{\begingroup \@sanitize@url \@href}%
	\providecommand \@href[1]{\@@startlink{#1}\@@href}%
	\providecommand \@@href[1]{\endgroup#1\@@endlink}%
	\providecommand \@sanitize@url [0]{\catcode `\\12\catcode `\$12\catcode
		`\&12\catcode `\#12\catcode `\^12\catcode `\_12\catcode `\%12\relax}%
	\providecommand \@@startlink[1]{}%
	\providecommand \@@endlink[0]{}%
	\providecommand \url  [0]{\begingroup\@sanitize@url \@url }%
	\providecommand \@url [1]{\endgroup\@href {#1}{\urlprefix }}%
	\providecommand \urlprefix  [0]{URL }%
	\providecommand \Eprint [0]{\href }%
	\providecommand \doibase [0]{https://doi.org/}%
	\providecommand \selectlanguage [0]{\@gobble}%
	\providecommand \bibinfo  [0]{\@secondoftwo}%
	\providecommand \bibfield  [0]{\@secondoftwo}%
	\providecommand \translation [1]{[#1]}%
	\providecommand \BibitemOpen [0]{}%
	\providecommand \bibitemStop [0]{}%
	\providecommand \bibitemNoStop [0]{.\EOS\space}%
	\providecommand \EOS [0]{\spacefactor3000\relax}%
	\providecommand \BibitemShut  [1]{\csname bibitem#1\endcsname}%
	\let\auto@bib@innerbib\@empty
	%</preamble>
	\bibitem [{\citenamefont {Dicke}(1954)}]{dicke1954coherence}%
	\BibitemOpen
	\bibfield  {author} {\bibinfo {author} {\bibfnamefont {R.~H.}\ \bibnamefont
			{Dicke}},\ }\bibfield  {title} {\bibinfo {title} {Coherence in spontaneous
			radiation processes},\ }\href
	{https://doi.org/https://doi.org/10.1103/PhysRev.93.99} {\bibfield  {journal}
		{\bibinfo  {journal} {Physical Review}\ }\textbf {\bibinfo {volume} {93}},\
		\bibinfo {pages} {99} (\bibinfo {year} {1954})}\BibitemShut {NoStop}%
	\bibitem [{\citenamefont {Rehler}\ and\ \citenamefont
		{Eberly}(1971)}]{rehler1971superradiance}%
	\BibitemOpen
	\bibfield  {author} {\bibinfo {author} {\bibfnamefont {N.~E.}\ \bibnamefont
			{Rehler}}\ and\ \bibinfo {author} {\bibfnamefont {J.~H.}\ \bibnamefont
			{Eberly}},\ }\bibfield  {title} {\bibinfo {title} {Superradiance},\ }\href
	{https://doi.org/https://doi.org/10.1103/PhysRevA.3.1735} {\bibfield
		{journal} {\bibinfo  {journal} {Physical Review A}\ }\textbf {\bibinfo
			{volume} {3}},\ \bibinfo {pages} {1735} (\bibinfo {year} {1971})}\BibitemShut
	{NoStop}%
	\bibitem [{\citenamefont {Gross}\ and\ \citenamefont
		{Haroche}(1982)}]{gross1982superradiance}%
	\BibitemOpen
	\bibfield  {author} {\bibinfo {author} {\bibfnamefont {M.}~\bibnamefont
			{Gross}}\ and\ \bibinfo {author} {\bibfnamefont {S.}~\bibnamefont
			{Haroche}},\ }\bibfield  {title} {\bibinfo {title} {Superradiance: An essay
			on the theory of collective spontaneous emission},\ }\href
	{https://doi.org/https://doi.org/10.1016/0370-1573(82)90102-8} {\bibfield
		{journal} {\bibinfo  {journal} {Physics Reports}\ }\textbf {\bibinfo {volume}
			{93}},\ \bibinfo {pages} {301} (\bibinfo {year} {1982})}\BibitemShut
	{NoStop}%
	\bibitem [{\citenamefont {Ams{\"u}ss}\ \emph {et~al.}(2011)\citenamefont
		{Ams{\"u}ss}, \citenamefont {Koller}, \citenamefont {N{\"o}bauer},
		\citenamefont {Putz}, \citenamefont {Rotter}, \citenamefont {Sandner},
		\citenamefont {Schneider}, \citenamefont {Schramb{\"o}ck}, \citenamefont
		{Steinhauser}, \citenamefont {Ritsch} \emph {et~al.}}]{amsuss2011cavity}%
	\BibitemOpen
	\bibfield  {author} {\bibinfo {author} {\bibfnamefont {R.}~\bibnamefont
			{Ams{\"u}ss}}, \bibinfo {author} {\bibfnamefont {C.}~\bibnamefont {Koller}},
		\bibinfo {author} {\bibfnamefont {T.}~\bibnamefont {N{\"o}bauer}}, \bibinfo
		{author} {\bibfnamefont {S.}~\bibnamefont {Putz}}, \bibinfo {author}
		{\bibfnamefont {S.}~\bibnamefont {Rotter}}, \bibinfo {author} {\bibfnamefont
			{K.}~\bibnamefont {Sandner}}, \bibinfo {author} {\bibfnamefont
			{S.}~\bibnamefont {Schneider}}, \bibinfo {author} {\bibfnamefont
			{M.}~\bibnamefont {Schramb{\"o}ck}}, \bibinfo {author} {\bibfnamefont
			{G.}~\bibnamefont {Steinhauser}}, \bibinfo {author} {\bibfnamefont
			{H.}~\bibnamefont {Ritsch}}, \emph {et~al.},\ }\bibfield  {title} {\bibinfo
		{title} {Cavity {QED} with magnetically coupled collective spin states},\
	}\href {https://doi.org/https://doi.org/10.1103/PhysRevLetters107.060502}
	{\bibfield  {journal} {\bibinfo  {journal} {Physical Review Letters}\
		}\textbf {\bibinfo {volume} {107}},\ \bibinfo {pages} {060502} (\bibinfo
		{year} {2011})}\BibitemShut {NoStop}%
	\bibitem [{\citenamefont {Sheremet}\ \emph {et~al.}(2023)\citenamefont
		{Sheremet}, \citenamefont {Petrov}, \citenamefont {Iorsh}, \citenamefont
		{Poshakinskiy},\ and\ \citenamefont {Poddubny}}]{sheremet2023waveguide}%
	\BibitemOpen
	\bibfield  {author} {\bibinfo {author} {\bibfnamefont {A.~S.}\ \bibnamefont
			{Sheremet}}, \bibinfo {author} {\bibfnamefont {M.~I.}\ \bibnamefont
			{Petrov}}, \bibinfo {author} {\bibfnamefont {I.~V.}\ \bibnamefont {Iorsh}},
		\bibinfo {author} {\bibfnamefont {A.~V.}\ \bibnamefont {Poshakinskiy}},\ and\
		\bibinfo {author} {\bibfnamefont {A.~N.}\ \bibnamefont {Poddubny}},\
	}\bibfield  {title} {\bibinfo {title} {Waveguide quantum electrodynamics:
			Collective radiance and photon-photon correlations},\ }\href
	{https://doi.org/https://doi.org/10.1103/RevModPhys.95.015002} {\bibfield
		{journal} {\bibinfo  {journal} {Reviews of Modern Physics}\ }\textbf
		{\bibinfo {volume} {95}},\ \bibinfo {pages} {015002} (\bibinfo {year}
		{2023})}\BibitemShut {NoStop}%
	\bibitem [{\citenamefont {Hotter}\ \emph {et~al.}(2024)\citenamefont {Hotter},
		\citenamefont {Ritsch},\ and\ \citenamefont {Gietka}}]{hotter2024combining}%
	\BibitemOpen
	\bibfield  {author} {\bibinfo {author} {\bibfnamefont {C.}~\bibnamefont
			{Hotter}}, \bibinfo {author} {\bibfnamefont {H.}~\bibnamefont {Ritsch}},\
		and\ \bibinfo {author} {\bibfnamefont {K.}~\bibnamefont {Gietka}},\
	}\bibfield  {title} {\bibinfo {title} {Combining critical and quantum
			metrology},\ }\href
	{https://doi.org/https://doi.org/10.1103/PhysRevLetters132.060801} {\bibfield
		{journal} {\bibinfo  {journal} {Physical Review Letters}\ }\textbf {\bibinfo
			{volume} {132}},\ \bibinfo {pages} {060801} (\bibinfo {year}
		{2024})}\BibitemShut {NoStop}%
	\bibitem [{\citenamefont {Saleem}\ \emph {et~al.}(2024)\citenamefont {Saleem},
		\citenamefont {Perlin}, \citenamefont {Shaji},\ and\ \citenamefont
		{Gray}}]{saleem2024achieving}%
	\BibitemOpen
	\bibfield  {author} {\bibinfo {author} {\bibfnamefont {Z.~H.}\ \bibnamefont
			{Saleem}}, \bibinfo {author} {\bibfnamefont {M.}~\bibnamefont {Perlin}},
		\bibinfo {author} {\bibfnamefont {A.}~\bibnamefont {Shaji}},\ and\ \bibinfo
		{author} {\bibfnamefont {S.~K.}\ \bibnamefont {Gray}},\ }\bibfield  {title}
	{\bibinfo {title} {Achieving the heisenberg limit with {Dicke} states in
			noisy quantum metrology},\ }\href
	{https://doi.org/https://doi.org/10.1103/PhysRevA.109.052615} {\bibfield
		{journal} {\bibinfo  {journal} {Physical Review A}\ }\textbf {\bibinfo
			{volume} {109}},\ \bibinfo {pages} {052615} (\bibinfo {year}
		{2024})}\BibitemShut {NoStop}%
	\bibitem [{\citenamefont {Gubaydullin}\ \emph {et~al.}(2026)\citenamefont
		{Gubaydullin}, \citenamefont {Slepnev}, \citenamefont {Mironov},\ and\
		\citenamefont {Vinokur}}]{gubaydullin2026quantum}%
	\BibitemOpen
	\bibfield  {author} {\bibinfo {author} {\bibfnamefont {A.}~\bibnamefont
			{Gubaydullin}}, \bibinfo {author} {\bibfnamefont {V.}~\bibnamefont
			{Slepnev}}, \bibinfo {author} {\bibfnamefont {A.}~\bibnamefont {Mironov}},\
		and\ \bibinfo {author} {\bibfnamefont {V.}~\bibnamefont {Vinokur}},\
	}\bibfield  {title} {\bibinfo {title} {Quantum magnetometry with
			superconducting-qubit greenberger--horne--zeilinger and dicke probes},\
	}\bibfield  {journal} {\bibinfo  {journal} {Scientific Reports}\ }\href
	{https://doi.org/10.1038/s41598-026-64752-w} {10.1038/s41598-026-64752-w}
	(\bibinfo {year} {2026})\BibitemShut {NoStop}%
	\bibitem [{\citenamefont {Lohof}\ \emph {et~al.}(2023)\citenamefont {Lohof},
		\citenamefont {Schumayer}, \citenamefont {Hutchinson},\ and\ \citenamefont
		{Gies}}]{lohof2023signatures}%
	\BibitemOpen
	\bibfield  {author} {\bibinfo {author} {\bibfnamefont {F.}~\bibnamefont
			{Lohof}}, \bibinfo {author} {\bibfnamefont {D.}~\bibnamefont {Schumayer}},
		\bibinfo {author} {\bibfnamefont {D.~A.}\ \bibnamefont {Hutchinson}},\ and\
		\bibinfo {author} {\bibfnamefont {C.}~\bibnamefont {Gies}},\ }\bibfield
	{title} {\bibinfo {title} {Signatures of superradiance as a witness to
			multipartite entanglement},\ }\href
	{https://doi.org/https://doi.org/10.1103/PhysRevLetters131.063601} {\bibfield
		{journal} {\bibinfo  {journal} {Physical review letters}\ }\textbf {\bibinfo
			{volume} {131}},\ \bibinfo {pages} {063601} (\bibinfo {year}
		{2023})}\BibitemShut {NoStop}%
	\bibitem [{\citenamefont {L{\"u}cke}\ \emph {et~al.}(2014)\citenamefont
		{L{\"u}cke}, \citenamefont {Peise}, \citenamefont {Vitagliano}, \citenamefont
		{Arlt}, \citenamefont {Santos}, \citenamefont {T{\'o}th},\ and\ \citenamefont
		{Klempt}}]{lucke2014detecting}%
	\BibitemOpen
	\bibfield  {author} {\bibinfo {author} {\bibfnamefont {B.}~\bibnamefont
			{L{\"u}cke}}, \bibinfo {author} {\bibfnamefont {J.}~\bibnamefont {Peise}},
		\bibinfo {author} {\bibfnamefont {G.}~\bibnamefont {Vitagliano}}, \bibinfo
		{author} {\bibfnamefont {J.}~\bibnamefont {Arlt}}, \bibinfo {author}
		{\bibfnamefont {L.}~\bibnamefont {Santos}}, \bibinfo {author} {\bibfnamefont
			{G.}~\bibnamefont {T{\'o}th}},\ and\ \bibinfo {author} {\bibfnamefont
			{C.}~\bibnamefont {Klempt}},\ }\bibfield  {title} {\bibinfo {title}
		{Detecting multiparticle entanglement of {Dicke} states},\ }\href
	{https://doi.org/https://doi.org/10.1103/PhysRevLetters112.155304} {\bibfield
		{journal} {\bibinfo  {journal} {Physical Review Letters}\ }\textbf {\bibinfo
			{volume} {112}},\ \bibinfo {pages} {155304} (\bibinfo {year}
		{2014})}\BibitemShut {NoStop}%
	\bibitem [{\citenamefont {Chiuri}\ \emph {et~al.}(2012)\citenamefont {Chiuri},
		\citenamefont {Greganti}, \citenamefont {Paternostro}, \citenamefont
		{Vallone},\ and\ \citenamefont {Mataloni}}]{chiuri2012experimental}%
	\BibitemOpen
	\bibfield  {author} {\bibinfo {author} {\bibfnamefont {A.}~\bibnamefont
			{Chiuri}}, \bibinfo {author} {\bibfnamefont {C.}~\bibnamefont {Greganti}},
		\bibinfo {author} {\bibfnamefont {M.}~\bibnamefont {Paternostro}}, \bibinfo
		{author} {\bibfnamefont {G.}~\bibnamefont {Vallone}},\ and\ \bibinfo {author}
		{\bibfnamefont {P.}~\bibnamefont {Mataloni}},\ }\bibfield  {title} {\bibinfo
		{title} {Experimental quantum networking protocols via four-qubit
			hyperentangled {Dicke} states},\ }\href
	{https://doi.org/https://doi.org/10.1103/PhysRevLetters109.173604} {\bibfield
		{journal} {\bibinfo  {journal} {Physical Review Letters}\ }\textbf {\bibinfo
			{volume} {109}},\ \bibinfo {pages} {173604} (\bibinfo {year}
		{2012})}\BibitemShut {NoStop}%
	\bibitem [{\citenamefont {Cheng}\ \emph {et~al.}(2020)\citenamefont {Cheng},
		\citenamefont {Liu}, \citenamefont {Guo}, \citenamefont {Chen}, \citenamefont
		{Zhang},\ and\ \citenamefont {Zhai}}]{cheng2020realizing}%
	\BibitemOpen
	\bibfield  {author} {\bibinfo {author} {\bibfnamefont {Y.}~\bibnamefont
			{Cheng}}, \bibinfo {author} {\bibfnamefont {C.}~\bibnamefont {Liu}}, \bibinfo
		{author} {\bibfnamefont {J.}~\bibnamefont {Guo}}, \bibinfo {author}
		{\bibfnamefont {Y.}~\bibnamefont {Chen}}, \bibinfo {author} {\bibfnamefont
			{P.}~\bibnamefont {Zhang}},\ and\ \bibinfo {author} {\bibfnamefont
			{H.}~\bibnamefont {Zhai}},\ }\bibfield  {title} {\bibinfo {title} {Realizing
			the {Hayden-Preskill} protocol with coupled {Dicke} models},\ }\href
	{https://doi.org/https://doi.org/10.1103/PhysRevResearch.2.043024} {\bibfield
		{journal} {\bibinfo  {journal} {Physical Review Research}\ }\textbf
		{\bibinfo {volume} {2}},\ \bibinfo {pages} {043024} (\bibinfo {year}
		{2020})}\BibitemShut {NoStop}%
	\bibitem [{\citenamefont {D{\"u}r}(2001)}]{dur2001multipartite}%
	\BibitemOpen
	\bibfield  {author} {\bibinfo {author} {\bibfnamefont {W.}~\bibnamefont
			{D{\"u}r}},\ }\bibfield  {title} {\bibinfo {title} {Multipartite entanglement
			that is robust against disposal of particles},\ }\href
	{https://doi.org/https://doi.org/10.1103/PhysRevA.63.020303} {\bibfield
		{journal} {\bibinfo  {journal} {Physical Review A}\ }\textbf {\bibinfo
			{volume} {63}},\ \bibinfo {pages} {020303} (\bibinfo {year}
		{2001})}\BibitemShut {NoStop}%
	\bibitem [{\citenamefont {Neven}\ \emph {et~al.}(2018)\citenamefont {Neven},
		\citenamefont {Martin},\ and\ \citenamefont
		{Bastin}}]{neven2018entanglement}%
	\BibitemOpen
	\bibfield  {author} {\bibinfo {author} {\bibfnamefont {A.}~\bibnamefont
			{Neven}}, \bibinfo {author} {\bibfnamefont {J.}~\bibnamefont {Martin}},\ and\
		\bibinfo {author} {\bibfnamefont {T.}~\bibnamefont {Bastin}},\ }\bibfield
	{title} {\bibinfo {title} {Entanglement robustness against particle loss in
			multiqubit systems},\ }\href
	{https://doi.org/https://doi.org/10.1103/PhysRevA.98.062335} {\bibfield
		{journal} {\bibinfo  {journal} {Physical Review A}\ }\textbf {\bibinfo
			{volume} {98}},\ \bibinfo {pages} {062335} (\bibinfo {year}
		{2018})}\BibitemShut {NoStop}%
	\bibitem [{\citenamefont {Sen}\ \emph {et~al.}(2003)\citenamefont {Sen},
		\citenamefont {Sen}, \citenamefont {Wie{\'s}niak}, \citenamefont
		{Kaszlikowski}, \citenamefont {{\.Z}ukowski} \emph
		{et~al.}}]{sen2003multiqubit}%
	\BibitemOpen
	\bibfield  {author} {\bibinfo {author} {\bibfnamefont {A.}~\bibnamefont
			{Sen}}, \bibinfo {author} {\bibfnamefont {U.}~\bibnamefont {Sen}}, \bibinfo
		{author} {\bibfnamefont {M.}~\bibnamefont {Wie{\'s}niak}}, \bibinfo {author}
		{\bibfnamefont {D.}~\bibnamefont {Kaszlikowski}}, \bibinfo {author}
		{\bibfnamefont {M.}~\bibnamefont {{\.Z}ukowski}}, \emph {et~al.},\ }\bibfield
	{title} {\bibinfo {title} {Multiqubit {W} states lead to stronger
			nonclassicality than {Greenberger-Horne-Zeilinger} states},\ }\href
	{https://doi.org/https://doi.org/10.1103/PhysRevA.68.062306} {\bibfield
		{journal} {\bibinfo  {journal} {Physical Review A}\ }\textbf {\bibinfo
			{volume} {68}},\ \bibinfo {pages} {062306} (\bibinfo {year}
		{2003})}\BibitemShut {NoStop}%
	\bibitem [{\citenamefont {Chen}\ \emph {et~al.}(2016)\citenamefont {Chen},
		\citenamefont {Ji}, \citenamefont {Yu},\ and\ \citenamefont
		{Zeng}}]{chen2016entanglement}%
	\BibitemOpen
	\bibfield  {author} {\bibinfo {author} {\bibfnamefont {J.-Y.}\ \bibnamefont
			{Chen}}, \bibinfo {author} {\bibfnamefont {Z.}~\bibnamefont {Ji}}, \bibinfo
		{author} {\bibfnamefont {N.}~\bibnamefont {Yu}},\ and\ \bibinfo {author}
		{\bibfnamefont {B.}~\bibnamefont {Zeng}},\ }\bibfield  {title} {\bibinfo
		{title} {Entanglement depth for symmetric states},\ }\href
	{https://doi.org/https://doi.org/10.1103/PhysRevA.94.042333} {\bibfield
		{journal} {\bibinfo  {journal} {Physical Review A}\ }\textbf {\bibinfo
			{volume} {94}},\ \bibinfo {pages} {042333} (\bibinfo {year}
		{2016})}\BibitemShut {NoStop}%
	\bibitem [{\citenamefont {Bhattacharyya}\ and\ \citenamefont
		{Roy}(2025)}]{bhattacharyya2025entanglement}%
	\BibitemOpen
	\bibfield  {author} {\bibinfo {author} {\bibfnamefont {S.}~\bibnamefont
			{Bhattacharyya}}\ and\ \bibinfo {author} {\bibfnamefont {S.}~\bibnamefont
			{Roy}},\ }\bibfield  {title} {\bibinfo {title} {Entanglement, coherence, and
			recursive linking in {Dicke} states: A topological perspective},\ }\bibfield
	{journal} {\bibinfo  {journal} {arXiv preprint arXiv:2512.12704}\ }\href
	{https://doi.org/https://doi.org/10.48550/arXiv.2512.12704}
	{https://doi.org/10.48550/arXiv.2512.12704} (\bibinfo {year}
	{2025})\BibitemShut {NoStop}%
	\bibitem [{\citenamefont {Bhattacharyya}\ \emph {et~al.}(2026)\citenamefont
		{Bhattacharyya}, \citenamefont {Ozaydin},\ and\ \citenamefont
		{Roy}}]{bhattacharyya2026super}%
	\BibitemOpen
	\bibfield  {author} {\bibinfo {author} {\bibfnamefont {S.}~\bibnamefont
			{Bhattacharyya}}, \bibinfo {author} {\bibfnamefont {F.}~\bibnamefont
			{Ozaydin}},\ and\ \bibinfo {author} {\bibfnamefont {S.}~\bibnamefont {Roy}},\
	}\bibfield  {title} {\bibinfo {title} {Super-link fragility in asymmetric
			$w$-class states under quantum noise},\ }\bibfield  {journal} {\bibinfo
		{journal} {arXiv preprint arXiv:2606.12307}\ }\href
	{https://doi.org/https://doi.org/10.48550/arXiv.2606.12307}
	{https://doi.org/10.48550/arXiv.2606.12307} (\bibinfo {year}
	{2026})\BibitemShut {NoStop}%
	\bibitem [{\citenamefont {Tashima}\ \emph {et~al.}(2008)\citenamefont
		{Tashima}, \citenamefont {{\"O}zdemir}, \citenamefont {Yamamoto},
		\citenamefont {Koashi},\ and\ \citenamefont {Imoto}}]{tashima2008elementary}%
	\BibitemOpen
	\bibfield  {author} {\bibinfo {author} {\bibfnamefont {T.}~\bibnamefont
			{Tashima}}, \bibinfo {author} {\bibfnamefont {{\c{S}}.~K.}\ \bibnamefont
			{{\"O}zdemir}}, \bibinfo {author} {\bibfnamefont {T.}~\bibnamefont
			{Yamamoto}}, \bibinfo {author} {\bibfnamefont {M.}~\bibnamefont {Koashi}},\
		and\ \bibinfo {author} {\bibfnamefont {N.}~\bibnamefont {Imoto}},\ }\bibfield
	{title} {\bibinfo {title} {Elementary optical gate for expanding an
			entanglement web},\ }\href {https://doi.org/10.1103/PhysRevA.77.030302}
	{\bibfield  {journal} {\bibinfo  {journal} {Physical Review A—Atomic,
				Molecular, and Optical Physics}\ }\textbf {\bibinfo {volume} {77}},\ \bibinfo
		{pages} {030302} (\bibinfo {year} {2008})}\BibitemShut {NoStop}%
	\bibitem [{\citenamefont {Tashima}\ \emph
		{et~al.}(2009{\natexlab{a}})\citenamefont {Tashima}, \citenamefont
		{{\"O}zdemir}, \citenamefont {Yamamoto}, \citenamefont {Koashi},\ and\
		\citenamefont {Imoto}}]{tashima2009local}%
	\BibitemOpen
	\bibfield  {author} {\bibinfo {author} {\bibfnamefont {T.}~\bibnamefont
			{Tashima}}, \bibinfo {author} {\bibfnamefont {{\c{S}}.~K.}\ \bibnamefont
			{{\"O}zdemir}}, \bibinfo {author} {\bibfnamefont {T.}~\bibnamefont
			{Yamamoto}}, \bibinfo {author} {\bibfnamefont {M.}~\bibnamefont {Koashi}},\
		and\ \bibinfo {author} {\bibfnamefont {N.}~\bibnamefont {Imoto}},\ }\bibfield
	{title} {\bibinfo {title} {Local expansion of photonic w state using a
			polarization-dependent beamsplitter},\ }\href
	{https://doi.org/10.1088/1367-2630/11/2/023024} {\bibfield  {journal}
		{\bibinfo  {journal} {New Journal of Physics}\ }\textbf {\bibinfo {volume}
			{11}},\ \bibinfo {pages} {023024} (\bibinfo {year}
		{2009}{\natexlab{a}})}\BibitemShut {NoStop}%
	\bibitem [{\citenamefont {Tashima}\ \emph
		{et~al.}(2009{\natexlab{b}})\citenamefont {Tashima}, \citenamefont
		{Wakatsuki}, \citenamefont {{\"O}zdemir}, \citenamefont {Yamamoto},
		\citenamefont {Koashi},\ and\ \citenamefont {Imoto}}]{tashima2009localPRL}%
	\BibitemOpen
	\bibfield  {author} {\bibinfo {author} {\bibfnamefont {T.}~\bibnamefont
			{Tashima}}, \bibinfo {author} {\bibfnamefont {T.}~\bibnamefont {Wakatsuki}},
		\bibinfo {author} {\bibfnamefont {{\c{S}}.~K.}\ \bibnamefont {{\"O}zdemir}},
		\bibinfo {author} {\bibfnamefont {T.}~\bibnamefont {Yamamoto}}, \bibinfo
		{author} {\bibfnamefont {M.}~\bibnamefont {Koashi}},\ and\ \bibinfo {author}
		{\bibfnamefont {N.}~\bibnamefont {Imoto}},\ }\bibfield  {title} {\bibinfo
		{title} {Local transformation of two einstein-podolsky-rosen photon pairs
			into a three-photon w state},\ }\href
	{https://doi.org/10.1103/PhysRevLett.102.130502} {\bibfield  {journal}
		{\bibinfo  {journal} {Physical Review Letters}\ }\textbf {\bibinfo {volume}
			{102}},\ \bibinfo {pages} {130502} (\bibinfo {year}
		{2009}{\natexlab{b}})}\BibitemShut {NoStop}%
	\bibitem [{\citenamefont {Tashima}\ \emph {et~al.}(2010)\citenamefont
		{Tashima}, \citenamefont {Kitano}, \citenamefont {{\"O}zdemir}, \citenamefont
		{Yamamoto}, \citenamefont {Koashi},\ and\ \citenamefont
		{Imoto}}]{tashima2010demonstration}%
	\BibitemOpen
	\bibfield  {author} {\bibinfo {author} {\bibfnamefont {T.}~\bibnamefont
			{Tashima}}, \bibinfo {author} {\bibfnamefont {T.}~\bibnamefont {Kitano}},
		\bibinfo {author} {\bibfnamefont {{\c{S}}.~K.}\ \bibnamefont {{\"O}zdemir}},
		\bibinfo {author} {\bibfnamefont {T.}~\bibnamefont {Yamamoto}}, \bibinfo
		{author} {\bibfnamefont {M.}~\bibnamefont {Koashi}},\ and\ \bibinfo {author}
		{\bibfnamefont {N.}~\bibnamefont {Imoto}},\ }\bibfield  {title} {\bibinfo
		{title} {Demonstration of local expansion toward large-scale entangled
			webs},\ }\href {https://doi.org/10.1103/PhysRevLett.105.210503} {\bibfield
		{journal} {\bibinfo  {journal} {Physical Review Letters}\ }\textbf {\bibinfo
			{volume} {105}},\ \bibinfo {pages} {210503} (\bibinfo {year}
		{2010})}\BibitemShut {NoStop}%
	\bibitem [{\citenamefont {Buğu}\ \emph {et~al.}(2013)\citenamefont {Buğu},
		\citenamefont {Yeşilyurt},\ and\ \citenamefont
		{Ozaydin}}]{bugu2013enhancing}%
	\BibitemOpen
	\bibfield  {author} {\bibinfo {author} {\bibfnamefont {S.}~\bibnamefont
			{Buğu}}, \bibinfo {author} {\bibfnamefont {C.}~\bibnamefont {Yeşilyurt}},\
		and\ \bibinfo {author} {\bibfnamefont {F.}~\bibnamefont {Ozaydin}},\
	}\bibfield  {title} {\bibinfo {title} {Enhancing the {W}-state
			quantum-network-fusion process with a single {F}redkin gate},\ }\href
	{https://doi.org/https://doi.org/10.1103/PhysRevA.87.032331} {\bibfield
		{journal} {\bibinfo  {journal} {Physical Review A}\ }\textbf {\bibinfo
			{volume} {87}},\ \bibinfo {pages} {032331} (\bibinfo {year}
		{2013})}\BibitemShut {NoStop}%
	\bibitem [{\citenamefont {Yesilyurt}\ \emph {et~al.}(2013)\citenamefont
		{Yesilyurt}, \citenamefont {Bugu},\ and\ \citenamefont
		{Ozaydin}}]{yesilyurt2013optical}%
	\BibitemOpen
	\bibfield  {author} {\bibinfo {author} {\bibfnamefont {C.}~\bibnamefont
			{Yesilyurt}}, \bibinfo {author} {\bibfnamefont {S.}~\bibnamefont {Bugu}},\
		and\ \bibinfo {author} {\bibfnamefont {F.}~\bibnamefont {Ozaydin}},\
	}\bibfield  {title} {\bibinfo {title} {An optical gate for simultaneous
			fusion of four photonic {W} or {B}ell states},\ }\href
	{https://doi.org/https://doi.org/10.1007/s11128-013-0578-9} {\bibfield
		{journal} {\bibinfo  {journal} {Quantum Information Processing}\ }\textbf
		{\bibinfo {volume} {12}},\ \bibinfo {pages} {2965} (\bibinfo {year}
		{2013})}\BibitemShut {NoStop}%
	\bibitem [{\citenamefont {Ozaydin}\ \emph {et~al.}(2014)\citenamefont
		{Ozaydin}, \citenamefont {Bugu}, \citenamefont {Yesilyurt}, \citenamefont
		{Altintas}, \citenamefont {Tame},\ and\ \citenamefont
		{{\"O}zdemir}}]{ozaydin2014fusing}%
	\BibitemOpen
	\bibfield  {author} {\bibinfo {author} {\bibfnamefont {F.}~\bibnamefont
			{Ozaydin}}, \bibinfo {author} {\bibfnamefont {S.}~\bibnamefont {Bugu}},
		\bibinfo {author} {\bibfnamefont {C.}~\bibnamefont {Yesilyurt}}, \bibinfo
		{author} {\bibfnamefont {A.~A.}\ \bibnamefont {Altintas}}, \bibinfo {author}
		{\bibfnamefont {M.}~\bibnamefont {Tame}},\ and\ \bibinfo {author}
		{\bibfnamefont {{\c{S}}.~K.}\ \bibnamefont {{\"O}zdemir}},\ }\bibfield
	{title} {\bibinfo {title} {Fusing multiple {W} states simultaneously with a
			{Fredkin} gate},\ }\href
	{https://doi.org/https://doi.org/10.1103/PhysRevA.89.042311} {\bibfield
		{journal} {\bibinfo  {journal} {Physical Review A}\ }\textbf {\bibinfo
			{volume} {89}},\ \bibinfo {pages} {042311} (\bibinfo {year}
		{2014})}\BibitemShut {NoStop}%
	\bibitem [{\citenamefont {Zang}\ \emph {et~al.}(2015)\citenamefont {Zang},
		\citenamefont {Yang}, \citenamefont {Ozaydin}, \citenamefont {Song},\ and\
		\citenamefont {Cao}}]{zang2015generating}%
	\BibitemOpen
	\bibfield  {author} {\bibinfo {author} {\bibfnamefont {X.-P.}\ \bibnamefont
			{Zang}}, \bibinfo {author} {\bibfnamefont {M.}~\bibnamefont {Yang}}, \bibinfo
		{author} {\bibfnamefont {F.}~\bibnamefont {Ozaydin}}, \bibinfo {author}
		{\bibfnamefont {W.}~\bibnamefont {Song}},\ and\ \bibinfo {author}
		{\bibfnamefont {Z.-L.}\ \bibnamefont {Cao}},\ }\bibfield  {title} {\bibinfo
		{title} {Generating multi-atom entangled {W} states via light-matter
			interface based fusion mechanism},\ }\href
	{https://doi.org/https://doi.org/10.1038/srep16245} {\bibfield  {journal}
		{\bibinfo  {journal} {Scientific Reports}\ }\textbf {\bibinfo {volume} {5}},\
		\bibinfo {pages} {16245} (\bibinfo {year} {2015})}\BibitemShut {NoStop}%
	\bibitem [{\citenamefont {Li}\ \emph {et~al.}(2016)\citenamefont {Li},
		\citenamefont {Kong}, \citenamefont {Yang}, \citenamefont {Ozaydin},
		\citenamefont {Yang},\ and\ \citenamefont {Cao}}]{li2016generating}%
	\BibitemOpen
	\bibfield  {author} {\bibinfo {author} {\bibfnamefont {K.}~\bibnamefont
			{Li}}, \bibinfo {author} {\bibfnamefont {F.~Z.}\ \bibnamefont {Kong}},
		\bibinfo {author} {\bibfnamefont {M.}~\bibnamefont {Yang}}, \bibinfo {author}
		{\bibfnamefont {F.}~\bibnamefont {Ozaydin}}, \bibinfo {author} {\bibfnamefont
			{Q.}~\bibnamefont {Yang}},\ and\ \bibinfo {author} {\bibfnamefont {Z.~L.}\
			\bibnamefont {Cao}},\ }\bibfield  {title} {\bibinfo {title} {Generating
			multi-photon {W}-like states for perfect quantum teleportation and superdense
			coding},\ }\href {https://doi.org/https://doi.org/10.1007/s11128-016-1332-x}
	{\bibfield  {journal} {\bibinfo  {journal} {Quantum Information Processing}\
		}\textbf {\bibinfo {volume} {15}},\ \bibinfo {pages} {3137} (\bibinfo {year}
		{2016})}\BibitemShut {NoStop}%
	\bibitem [{\citenamefont {Al~Farooqui}\ \emph {et~al.}(2015)\citenamefont
		{Al~Farooqui}, \citenamefont {Breeland}, \citenamefont {Aslam}, \citenamefont
		{Sadatgol}, \citenamefont {{\"O}zdemir}, \citenamefont {Tame}, \citenamefont
		{Yang},\ and\ \citenamefont {G{\"u}ney}}]{al2015quantum}%
	\BibitemOpen
	\bibfield  {author} {\bibinfo {author} {\bibfnamefont {M.~A.}\ \bibnamefont
			{Al~Farooqui}}, \bibinfo {author} {\bibfnamefont {J.}~\bibnamefont
			{Breeland}}, \bibinfo {author} {\bibfnamefont {M.~I.}\ \bibnamefont {Aslam}},
		\bibinfo {author} {\bibfnamefont {M.}~\bibnamefont {Sadatgol}}, \bibinfo
		{author} {\bibfnamefont {{\c{S}}.~K.}\ \bibnamefont {{\"O}zdemir}}, \bibinfo
		{author} {\bibfnamefont {M.}~\bibnamefont {Tame}}, \bibinfo {author}
		{\bibfnamefont {L.}~\bibnamefont {Yang}},\ and\ \bibinfo {author}
		{\bibfnamefont {D.~{\"O}.}\ \bibnamefont {G{\"u}ney}},\ }\bibfield  {title}
	{\bibinfo {title} {Quantum entanglement distillation with metamaterials},\
	}\href {https://doi.org/10.1364/OE.23.017941} {\bibfield  {journal} {\bibinfo
			{journal} {Optics express}\ }\textbf {\bibinfo {volume} {23}},\ \bibinfo
		{pages} {17941} (\bibinfo {year} {2015})}\BibitemShut {NoStop}%
	\bibitem [{\citenamefont {Yesilyurt}\ \emph {et~al.}(2016)\citenamefont
		{Yesilyurt}, \citenamefont {Bugu}, \citenamefont {Ozaydin}, \citenamefont
		{Altintas}, \citenamefont {Tame}, \citenamefont {Yang},\ and\ \citenamefont
		{{\"O}zdemir}}]{yesilyurt2016deterministic}%
	\BibitemOpen
	\bibfield  {author} {\bibinfo {author} {\bibfnamefont {C.}~\bibnamefont
			{Yesilyurt}}, \bibinfo {author} {\bibfnamefont {S.}~\bibnamefont {Bugu}},
		\bibinfo {author} {\bibfnamefont {F.}~\bibnamefont {Ozaydin}}, \bibinfo
		{author} {\bibfnamefont {A.~A.}\ \bibnamefont {Altintas}}, \bibinfo {author}
		{\bibfnamefont {M.}~\bibnamefont {Tame}}, \bibinfo {author} {\bibfnamefont
			{L.}~\bibnamefont {Yang}},\ and\ \bibinfo {author} {\bibfnamefont
			{{\c{S}}.~K.}\ \bibnamefont {{\"O}zdemir}},\ }\bibfield  {title} {\bibinfo
		{title} {Deterministic local doubling of {W} states},\ }\href
	{https://doi.org/https://doi.org/10.1364/JOSAB.33.002313} {\bibfield
		{journal} {\bibinfo  {journal} {Journal of the Optical Society of America B}\
		}\textbf {\bibinfo {volume} {33}},\ \bibinfo {pages} {2313} (\bibinfo {year}
		{2016})}\BibitemShut {NoStop}%
	\bibitem [{\citenamefont {Zang}\ \emph {et~al.}(2016)\citenamefont {Zang},
		\citenamefont {Yang}, \citenamefont {Ozaydin}, \citenamefont {Song},\ and\
		\citenamefont {Cao}}]{zang2016deterministic}%
	\BibitemOpen
	\bibfield  {author} {\bibinfo {author} {\bibfnamefont {X.-P.}\ \bibnamefont
			{Zang}}, \bibinfo {author} {\bibfnamefont {M.}~\bibnamefont {Yang}}, \bibinfo
		{author} {\bibfnamefont {F.}~\bibnamefont {Ozaydin}}, \bibinfo {author}
		{\bibfnamefont {W.}~\bibnamefont {Song}},\ and\ \bibinfo {author}
		{\bibfnamefont {Z.-L.}\ \bibnamefont {Cao}},\ }\bibfield  {title} {\bibinfo
		{title} {Deterministic generation of large scale atomic {W} states},\ }\href
	{https://doi.org/https://doi.org/10.1364/OE.24.012293} {\bibfield  {journal}
		{\bibinfo  {journal} {Optics Express}\ }\textbf {\bibinfo {volume} {24}},\
		\bibinfo {pages} {12293} (\bibinfo {year} {2016})}\BibitemShut {NoStop}%
	\bibitem [{\citenamefont {Ozaydin}\ \emph {et~al.}(2021)\citenamefont
		{Ozaydin}, \citenamefont {Yesilyurt}, \citenamefont {Bugu},\ and\
		\citenamefont {Koashi}}]{ozaydin2021deterministic}%
	\BibitemOpen
	\bibfield  {author} {\bibinfo {author} {\bibfnamefont {F.}~\bibnamefont
			{Ozaydin}}, \bibinfo {author} {\bibfnamefont {C.}~\bibnamefont {Yesilyurt}},
		\bibinfo {author} {\bibfnamefont {S.}~\bibnamefont {Bugu}},\ and\ \bibinfo
		{author} {\bibfnamefont {M.}~\bibnamefont {Koashi}},\ }\bibfield  {title}
	{\bibinfo {title} {Deterministic preparation of {$W$} states via spin-photon
			interactions},\ }\href
	{https://doi.org/https://doi.org/10.1103/PhysRevA.103.052421} {\bibfield
		{journal} {\bibinfo  {journal} {Physical Review A}\ }\textbf {\bibinfo
			{volume} {103}},\ \bibinfo {pages} {052421} (\bibinfo {year}
		{2021})}\BibitemShut {NoStop}%
	\bibitem [{\citenamefont {Bugu}\ \emph {et~al.}(2020)\citenamefont {Bugu},
		\citenamefont {Ozaydin}, \citenamefont {Ferrus},\ and\ \citenamefont
		{Kodera}}]{bugu2020preparing}%
	\BibitemOpen
	\bibfield  {author} {\bibinfo {author} {\bibfnamefont {S.}~\bibnamefont
			{Bugu}}, \bibinfo {author} {\bibfnamefont {F.}~\bibnamefont {Ozaydin}},
		\bibinfo {author} {\bibfnamefont {T.}~\bibnamefont {Ferrus}},\ and\ \bibinfo
		{author} {\bibfnamefont {T.}~\bibnamefont {Kodera}},\ }\bibfield  {title}
	{\bibinfo {title} {Preparing multipartite entangled spin qubits via {Pauli}
			spin blockade},\ }\href
	{https://doi.org/https://doi.org/10.1038/s41598-020-60299-6} {\bibfield
		{journal} {\bibinfo  {journal} {Scientific Reports}\ }\textbf {\bibinfo
			{volume} {10}},\ \bibinfo {pages} {3481} (\bibinfo {year}
		{2020})}\BibitemShut {NoStop}%
	\bibitem [{\citenamefont {Kim}\ \emph {et~al.}(2020)\citenamefont {Kim},
		\citenamefont {Cho}, \citenamefont {Lim},\ and\ \citenamefont
		{Han}}]{kim2020efficient}%
	\BibitemOpen
	\bibfield  {author} {\bibinfo {author} {\bibfnamefont {Y.-S.}\ \bibnamefont
			{Kim}}, \bibinfo {author} {\bibfnamefont {Y.-W.}\ \bibnamefont {Cho}},
		\bibinfo {author} {\bibfnamefont {H.-T.}\ \bibnamefont {Lim}},\ and\ \bibinfo
		{author} {\bibfnamefont {S.-W.}\ \bibnamefont {Han}},\ }\bibfield  {title}
	{\bibinfo {title} {Efficient linear optical generation of a multipartite w
			state via a quantum eraser},\ }\href
	{https://doi.org/https://doi.org/10.1103/PhysRevA.101.022337} {\bibfield
		{journal} {\bibinfo  {journal} {Physical Review A}\ }\textbf {\bibinfo
			{volume} {101}},\ \bibinfo {pages} {022337} (\bibinfo {year}
		{2020})}\BibitemShut {NoStop}%
	\bibitem [{\citenamefont {Chakraborty}\ \emph {et~al.}(2014)\citenamefont
		{Chakraborty}, \citenamefont {Choi}, \citenamefont {Maitra},\ and\
		\citenamefont {Maitra}}]{chakraborty2014efficient}%
	\BibitemOpen
	\bibfield  {author} {\bibinfo {author} {\bibfnamefont {K.}~\bibnamefont
			{Chakraborty}}, \bibinfo {author} {\bibfnamefont {B.-S.}\ \bibnamefont
			{Choi}}, \bibinfo {author} {\bibfnamefont {A.}~\bibnamefont {Maitra}},\ and\
		\bibinfo {author} {\bibfnamefont {S.}~\bibnamefont {Maitra}},\ }\bibfield
	{title} {\bibinfo {title} {Efficient quantum algorithms to construct
			arbitrary {Dicke} states},\ }\href
	{https://doi.org/https://doi.org/10.1007/s11128-014-0797-8} {\bibfield
		{journal} {\bibinfo  {journal} {Quantum information processing}\ }\textbf
		{\bibinfo {volume} {13}},\ \bibinfo {pages} {2049} (\bibinfo {year}
		{2014})}\BibitemShut {NoStop}%
	\bibitem [{\citenamefont {B{\"a}rtschi}\ and\ \citenamefont
		{Eidenbenz}(2019)}]{bartschi2019deterministic}%
	\BibitemOpen
	\bibfield  {author} {\bibinfo {author} {\bibfnamefont {A.}~\bibnamefont
			{B{\"a}rtschi}}\ and\ \bibinfo {author} {\bibfnamefont {S.}~\bibnamefont
			{Eidenbenz}},\ }\bibfield  {title} {\bibinfo {title} {Deterministic
			preparation of {Dicke} states},\ }in\ \href
	{https://doi.org/https://doi.org/10.1007/978-3-030-25027-0_9} {\emph
		{\bibinfo {booktitle} {International Symposium on Fundamentals of Computation
				Theory}}}\ (\bibinfo {organization} {Springer},\ \bibinfo {year} {2019})\
	pp.\ \bibinfo {pages} {126--139}\BibitemShut {NoStop}%
	\bibitem [{\citenamefont {Mukherjee}\ \emph {et~al.}(2020)\citenamefont
		{Mukherjee}, \citenamefont {Maitra}, \citenamefont {Gaurav},\ and\
		\citenamefont {Roy}}]{mukherjee2020preparing}%
	\BibitemOpen
	\bibfield  {author} {\bibinfo {author} {\bibfnamefont {C.~S.}\ \bibnamefont
			{Mukherjee}}, \bibinfo {author} {\bibfnamefont {S.}~\bibnamefont {Maitra}},
		\bibinfo {author} {\bibfnamefont {V.}~\bibnamefont {Gaurav}},\ and\ \bibinfo
		{author} {\bibfnamefont {D.}~\bibnamefont {Roy}},\ }\bibfield  {title}
	{\bibinfo {title} {Preparing {Dicke} states on a quantum computer},\ }\href
	{https://doi.org/https://doi.org/10.1109/TQE.2020.3041479} {\bibfield
		{journal} {\bibinfo  {journal} {IEEE Transactions on Quantum Engineering}\
		}\textbf {\bibinfo {volume} {1}},\ \bibinfo {pages} {1} (\bibinfo {year}
		{2020})}\BibitemShut {NoStop}%
	\bibitem [{\citenamefont {Aktar}\ \emph {et~al.}(2022)\citenamefont {Aktar},
		\citenamefont {B{\"a}rtschi}, \citenamefont {Badawy},\ and\ \citenamefont
		{Eidenbenz}}]{aktar2022divide}%
	\BibitemOpen
	\bibfield  {author} {\bibinfo {author} {\bibfnamefont {S.}~\bibnamefont
			{Aktar}}, \bibinfo {author} {\bibfnamefont {A.}~\bibnamefont {B{\"a}rtschi}},
		\bibinfo {author} {\bibfnamefont {A.-H.~A.}\ \bibnamefont {Badawy}},\ and\
		\bibinfo {author} {\bibfnamefont {S.}~\bibnamefont {Eidenbenz}},\ }\bibfield
	{title} {\bibinfo {title} {A divide-and-conquer approach to {Dicke} state
			preparation},\ }\href
	{https://doi.org/https://doi.org/10.1109/TQE.2022.3174547} {\bibfield
		{journal} {\bibinfo  {journal} {IEEE Transactions on Quantum Engineering}\
		}\textbf {\bibinfo {volume} {3}},\ \bibinfo {pages} {1} (\bibinfo {year}
		{2022})}\BibitemShut {NoStop}%
	\bibitem [{\citenamefont {Yu}\ \emph {et~al.}(2024)\citenamefont {Yu},
		\citenamefont {Muleady}, \citenamefont {Wang}, \citenamefont {Schine},
		\citenamefont {Gorshkov},\ and\ \citenamefont {Childs}}]{yu2024efficient}%
	\BibitemOpen
	\bibfield  {author} {\bibinfo {author} {\bibfnamefont {J.}~\bibnamefont
			{Yu}}, \bibinfo {author} {\bibfnamefont {S.~R.}\ \bibnamefont {Muleady}},
		\bibinfo {author} {\bibfnamefont {Y.-X.}\ \bibnamefont {Wang}}, \bibinfo
		{author} {\bibfnamefont {N.}~\bibnamefont {Schine}}, \bibinfo {author}
		{\bibfnamefont {A.~V.}\ \bibnamefont {Gorshkov}},\ and\ \bibinfo {author}
		{\bibfnamefont {A.~M.}\ \bibnamefont {Childs}},\ }\bibfield  {title}
	{\bibinfo {title} {Efficient preparation of {Dicke} states},\ }\bibfield
	{journal} {\bibinfo  {journal} {arXiv preprint arXiv:2411.03428}\ }\href
	{https://doi.org/https://doi.org/10.48550/arXiv.2411.03428}
	{https://doi.org/10.48550/arXiv.2411.03428} (\bibinfo {year}
	{2024})\BibitemShut {NoStop}%
	\bibitem [{\citenamefont {Stojanovi{\'c}}\ and\ \citenamefont
		{Nauth}(2023)}]{stojanovic2023dicke}%
	\BibitemOpen
	\bibfield  {author} {\bibinfo {author} {\bibfnamefont {V.~M.}\ \bibnamefont
			{Stojanovi{\'c}}}\ and\ \bibinfo {author} {\bibfnamefont {J.~K.}\
			\bibnamefont {Nauth}},\ }\bibfield  {title} {\bibinfo {title} {Dicke-state
			preparation through global transverse control of ising-coupled qubits},\
	}\href {https://doi.org/https://doi.org/10.1103/PhysRevA.108.012608}
	{\bibfield  {journal} {\bibinfo  {journal} {Physical Review A}\ }\textbf
		{\bibinfo {volume} {108}},\ \bibinfo {pages} {012608} (\bibinfo {year}
		{2023})}\BibitemShut {NoStop}%
	\bibitem [{\citenamefont {Wang}\ and\ \citenamefont
		{Terhal}(2021)}]{wang2021preparing}%
	\BibitemOpen
	\bibfield  {author} {\bibinfo {author} {\bibfnamefont {Y.}~\bibnamefont
			{Wang}}\ and\ \bibinfo {author} {\bibfnamefont {B.~M.}\ \bibnamefont
			{Terhal}},\ }\bibfield  {title} {\bibinfo {title} {Preparing {Dicke} states
			in a spin ensemble using phase estimation},\ }\href
	{https://doi.org/https://doi.org/10.1103/PhysRevA.104.032407} {\bibfield
		{journal} {\bibinfo  {journal} {Physical Review A}\ }\textbf {\bibinfo
			{volume} {104}},\ \bibinfo {pages} {032407} (\bibinfo {year}
		{2021})}\BibitemShut {NoStop}%
	\bibitem [{\citenamefont {Linington}\ and\ \citenamefont
		{Vitanov}(2008)}]{linington2008robust}%
	\BibitemOpen
	\bibfield  {author} {\bibinfo {author} {\bibfnamefont {I.}~\bibnamefont
			{Linington}}\ and\ \bibinfo {author} {\bibfnamefont {N.}~\bibnamefont
			{Vitanov}},\ }\bibfield  {title} {\bibinfo {title} {Robust creation of
			arbitrary-sized dicke states of trapped ions by global addressing},\ }\href
	{https://doi.org/https://doi.org/10.1103/PhysRevA.77.010302} {\bibfield
		{journal} {\bibinfo  {journal} {Physical Review A}\ }\textbf {\bibinfo
			{volume} {77}},\ \bibinfo {pages} {010302} (\bibinfo {year}
		{2008})}\BibitemShut {NoStop}%
	\bibitem [{\citenamefont {Opatrn{\`y}}\ \emph {et~al.}(2016)\citenamefont
		{Opatrn{\`y}}, \citenamefont {Saberi}, \citenamefont {Brion},\ and\
		\citenamefont {M{\o}lmer}}]{opatrny2016counterdiabatic}%
	\BibitemOpen
	\bibfield  {author} {\bibinfo {author} {\bibfnamefont {T.}~\bibnamefont
			{Opatrn{\`y}}}, \bibinfo {author} {\bibfnamefont {H.}~\bibnamefont {Saberi}},
		\bibinfo {author} {\bibfnamefont {E.}~\bibnamefont {Brion}},\ and\ \bibinfo
		{author} {\bibfnamefont {K.}~\bibnamefont {M{\o}lmer}},\ }\bibfield  {title}
	{\bibinfo {title} {Counterdiabatic driving in spin squeezing and
			{Dicke}-state preparation},\ }\href
	{https://doi.org/https://doi.org/10.1103/PhysRevA.93.023815} {\bibfield
		{journal} {\bibinfo  {journal} {Physical Review A}\ }\textbf {\bibinfo
			{volume} {93}},\ \bibinfo {pages} {023815} (\bibinfo {year}
		{2016})}\BibitemShut {NoStop}%
	\bibitem [{\citenamefont {Carrasco}\ \emph {et~al.}(2024)\citenamefont
		{Carrasco}, \citenamefont {Goerz}, \citenamefont {Malinovskaya},
		\citenamefont {Vuleti{\'c}}, \citenamefont {Schleich},\ and\ \citenamefont
		{Malinovsky}}]{carrasco2024dicke}%
	\BibitemOpen
	\bibfield  {author} {\bibinfo {author} {\bibfnamefont {S.~C.}\ \bibnamefont
			{Carrasco}}, \bibinfo {author} {\bibfnamefont {M.~H.}\ \bibnamefont {Goerz}},
		\bibinfo {author} {\bibfnamefont {S.~A.}\ \bibnamefont {Malinovskaya}},
		\bibinfo {author} {\bibfnamefont {V.}~\bibnamefont {Vuleti{\'c}}}, \bibinfo
		{author} {\bibfnamefont {W.~P.}\ \bibnamefont {Schleich}},\ and\ \bibinfo
		{author} {\bibfnamefont {V.~S.}\ \bibnamefont {Malinovsky}},\ }\bibfield
	{title} {\bibinfo {title} {Dicke state generation and extreme spin squeezing
			via rapid adiabatic passage},\ }\href
	{https://doi.org/https://doi.org/10.1103/PhysRevLetters132.153603} {\bibfield
		{journal} {\bibinfo  {journal} {Physical Review Letters}\ }\textbf {\bibinfo
			{volume} {132}},\ \bibinfo {pages} {153603} (\bibinfo {year}
		{2024})}\BibitemShut {NoStop}%
	\bibitem [{\citenamefont {Zhu}\ \emph {et~al.}(2025)\citenamefont {Zhu},
		\citenamefont {So}, \citenamefont {Pagano},\ and\ \citenamefont
		{Pu}}]{zhu2025dissipation}%
	\BibitemOpen
	\bibfield  {author} {\bibinfo {author} {\bibfnamefont {M.}~\bibnamefont
			{Zhu}}, \bibinfo {author} {\bibfnamefont {V.}~\bibnamefont {So}}, \bibinfo
		{author} {\bibfnamefont {G.}~\bibnamefont {Pagano}},\ and\ \bibinfo {author}
		{\bibfnamefont {H.}~\bibnamefont {Pu}},\ }\bibfield  {title} {\bibinfo
		{title} {Dissipation-assisted steady-state entanglement engineering based on
			electron transfer models},\ }\href
	{https://doi.org/https://doi.org/10.1103/9sr4-3jz2} {\bibfield  {journal}
		{\bibinfo  {journal} {Physical Review A}\ }\textbf {\bibinfo {volume}
			{112}},\ \bibinfo {pages} {012617} (\bibinfo {year} {2025})}\BibitemShut
	{NoStop}%
	\bibitem [{\citenamefont {Kobayashi}\ \emph {et~al.}(2014)\citenamefont
		{Kobayashi}, \citenamefont {Ikuta}, \citenamefont {{\"O}zdemir},
		\citenamefont {Tame}, \citenamefont {Yamamoto}, \citenamefont {Koashi},\ and\
		\citenamefont {Imoto}}]{kobayashi2014universal}%
	\BibitemOpen
	\bibfield  {author} {\bibinfo {author} {\bibfnamefont {T.}~\bibnamefont
			{Kobayashi}}, \bibinfo {author} {\bibfnamefont {R.}~\bibnamefont {Ikuta}},
		\bibinfo {author} {\bibfnamefont {{\c{S}}.~K.}\ \bibnamefont {{\"O}zdemir}},
		\bibinfo {author} {\bibfnamefont {M.}~\bibnamefont {Tame}}, \bibinfo {author}
		{\bibfnamefont {T.}~\bibnamefont {Yamamoto}}, \bibinfo {author}
		{\bibfnamefont {M.}~\bibnamefont {Koashi}},\ and\ \bibinfo {author}
		{\bibfnamefont {N.}~\bibnamefont {Imoto}},\ }\bibfield  {title} {\bibinfo
		{title} {Universal gates for transforming multipartite entangled {Dicke}
			states},\ }\href
	{https://doi.org/https://doi.org/10.1088/1367-2630/16/2/023005} {\bibfield
		{journal} {\bibinfo  {journal} {New Journal of Physics}\ }\textbf {\bibinfo
			{volume} {16}},\ \bibinfo {pages} {023005} (\bibinfo {year}
		{2014})}\BibitemShut {NoStop}%
	\bibitem [{\citenamefont {Thapa}\ \emph {et~al.}(2025)\citenamefont {Thapa},
		\citenamefont {Moran}, \citenamefont {Vu},\ and\ \citenamefont
		{Ozaydin}}]{thapa2025expanding}%
	\BibitemOpen
	\bibfield  {author} {\bibinfo {author} {\bibfnamefont {B.}~\bibnamefont
			{Thapa}}, \bibinfo {author} {\bibfnamefont {O.}~\bibnamefont {Moran}},
		\bibinfo {author} {\bibfnamefont {D.-K.}\ \bibnamefont {Vu}},\ and\ \bibinfo
		{author} {\bibfnamefont {F.}~\bibnamefont {Ozaydin}},\ }\bibfield  {title}
	{\bibinfo {title} {Expanding a 4-qubit {Dicke} state to a 5-qubit {Dicke}
			state with limited qubit access},\ }\href
	{https://doi.org/https://doi.org/10.1007/s11128-025-05021-z} {\bibfield
		{journal} {\bibinfo  {journal} {Quantum Information Processing}\ }\textbf
		{\bibinfo {volume} {24}},\ \bibinfo {pages} {401} (\bibinfo {year}
		{2025})}\BibitemShut {NoStop}%
	\bibitem [{\citenamefont {Wang}\ \emph {et~al.}(2026)\citenamefont {Wang},
		\citenamefont {Scully},\ and\ \citenamefont
		{Agarwal}}]{wang2026deterministic}%
	\BibitemOpen
	\bibfield  {author} {\bibinfo {author} {\bibfnamefont {H.}~\bibnamefont
			{Wang}}, \bibinfo {author} {\bibfnamefont {M.~O.}\ \bibnamefont {Scully}},\
		and\ \bibinfo {author} {\bibfnamefont {G.~S.}\ \bibnamefont {Agarwal}},\
	}\bibfield  {title} {\bibinfo {title} {Deterministic preparation of entangled
			dicke states},\ }\bibfield  {journal} {\bibinfo  {journal} {arXiv preprint
			arXiv:2608.22168}\ }\href {https://doi.org/10.48550/arXiv.2608.22168}
	{10.48550/arXiv.2608.22168} (\bibinfo {year} {2026})\BibitemShut {NoStop}%
	\bibitem [{\citenamefont {Preskill}(2018)}]{preskill2018quantum}%
	\BibitemOpen
	\bibfield  {author} {\bibinfo {author} {\bibfnamefont {J.}~\bibnamefont
			{Preskill}},\ }\bibfield  {title} {\bibinfo {title} {Quantum computing in the
			{NISQ} era and beyond},\ }\href
	{https://doi.org/https://doi.org/10.22331/q-2018-08-06-79} {\bibfield
		{journal} {\bibinfo  {journal} {Quantum}\ }\textbf {\bibinfo {volume} {2}},\
		\bibinfo {pages} {79} (\bibinfo {year} {2018})}\BibitemShut {NoStop}%
	\bibitem [{\citenamefont {Arute}\ \emph {et~al.}(2019)\citenamefont {Arute},
		\citenamefont {Arya}, \citenamefont {Babbush}, \citenamefont {Bacon},
		\citenamefont {Bardin}, \citenamefont {Barends}, \citenamefont {Biswas},
		\citenamefont {Boixo}, \citenamefont {Brandao}, \citenamefont {Buell} \emph
		{et~al.}}]{arute2019quantum}%
	\BibitemOpen
	\bibfield  {author} {\bibinfo {author} {\bibfnamefont {F.}~\bibnamefont
			{Arute}}, \bibinfo {author} {\bibfnamefont {K.}~\bibnamefont {Arya}},
		\bibinfo {author} {\bibfnamefont {R.}~\bibnamefont {Babbush}}, \bibinfo
		{author} {\bibfnamefont {D.}~\bibnamefont {Bacon}}, \bibinfo {author}
		{\bibfnamefont {J.~C.}\ \bibnamefont {Bardin}}, \bibinfo {author}
		{\bibfnamefont {R.}~\bibnamefont {Barends}}, \bibinfo {author} {\bibfnamefont
			{R.}~\bibnamefont {Biswas}}, \bibinfo {author} {\bibfnamefont
			{S.}~\bibnamefont {Boixo}}, \bibinfo {author} {\bibfnamefont {F.~G.}\
			\bibnamefont {Brandao}}, \bibinfo {author} {\bibfnamefont {D.~A.}\
			\bibnamefont {Buell}}, \emph {et~al.},\ }\bibfield  {title} {\bibinfo {title}
		{Quantum supremacy using a programmable superconducting processor},\ }\href
	{https://doi.org/https://doi.org/10.1038/s41586-019-1666-5} {\bibfield
		{journal} {\bibinfo  {journal} {Nature}\ }\textbf {\bibinfo {volume} {574}},\
		\bibinfo {pages} {505} (\bibinfo {year} {2019})}\BibitemShut {NoStop}%
	\bibitem [{\citenamefont {Temme}\ \emph {et~al.}(2017)\citenamefont {Temme},
		\citenamefont {Bravyi},\ and\ \citenamefont {Gambetta}}]{temme2017error}%
	\BibitemOpen
	\bibfield  {author} {\bibinfo {author} {\bibfnamefont {K.}~\bibnamefont
			{Temme}}, \bibinfo {author} {\bibfnamefont {S.}~\bibnamefont {Bravyi}},\ and\
		\bibinfo {author} {\bibfnamefont {J.~M.}\ \bibnamefont {Gambetta}},\
	}\bibfield  {title} {\bibinfo {title} {Error mitigation for short-depth
			quantum circuits},\ }\href
	{https://doi.org/https://doi.org/10.1103/PhysRevLetters119.180509} {\bibfield
		{journal} {\bibinfo  {journal} {Physical Review Letters}\ }\textbf {\bibinfo
			{volume} {119}},\ \bibinfo {pages} {180509} (\bibinfo {year}
		{2017})}\BibitemShut {NoStop}%
	\bibitem [{\citenamefont {Endo}\ \emph {et~al.}(2021)\citenamefont {Endo},
		\citenamefont {Cai}, \citenamefont {Benjamin},\ and\ \citenamefont
		{Yuan}}]{endo2021hybrid}%
	\BibitemOpen
	\bibfield  {author} {\bibinfo {author} {\bibfnamefont {S.}~\bibnamefont
			{Endo}}, \bibinfo {author} {\bibfnamefont {Z.}~\bibnamefont {Cai}}, \bibinfo
		{author} {\bibfnamefont {S.~C.}\ \bibnamefont {Benjamin}},\ and\ \bibinfo
		{author} {\bibfnamefont {X.}~\bibnamefont {Yuan}},\ }\bibfield  {title}
	{\bibinfo {title} {Hybrid quantum-classical algorithms and quantum error
			mitigation},\ }\href {https://doi.org/https://doi.org/10.7566/JPSJ.90.032001}
	{\bibfield  {journal} {\bibinfo  {journal} {Journal of the Physical Society
				of Japan}\ }\textbf {\bibinfo {volume} {90}},\ \bibinfo {pages} {032001}
		(\bibinfo {year} {2021})}\BibitemShut {NoStop}%
	\bibitem [{\citenamefont {Ziman}\ \emph {et~al.}(2002)\citenamefont {Ziman},
		\citenamefont {{\v{S}}telmachovi{\v{c}}}, \citenamefont {Bu{\v{z}}ek},
		\citenamefont {Hillery}, \citenamefont {Scarani},\ and\ \citenamefont
		{Gisin}}]{ziman2002diluting}%
	\BibitemOpen
	\bibfield  {author} {\bibinfo {author} {\bibfnamefont {M.}~\bibnamefont
			{Ziman}}, \bibinfo {author} {\bibfnamefont {P.}~\bibnamefont
			{{\v{S}}telmachovi{\v{c}}}}, \bibinfo {author} {\bibfnamefont
			{V.}~\bibnamefont {Bu{\v{z}}ek}}, \bibinfo {author} {\bibfnamefont
			{M.}~\bibnamefont {Hillery}}, \bibinfo {author} {\bibfnamefont
			{V.}~\bibnamefont {Scarani}},\ and\ \bibinfo {author} {\bibfnamefont
			{N.}~\bibnamefont {Gisin}},\ }\bibfield  {title} {\bibinfo {title} {Diluting
			quantum information: An analysis of information transfer in system-reservoir
			interactions},\ }\href {https://doi.org/10.1103/PhysRevA.65.042105}
	{\bibfield  {journal} {\bibinfo  {journal} {Physical Review A}\ }\textbf
		{\bibinfo {volume} {65}},\ \bibinfo {pages} {042105} (\bibinfo {year}
		{2002})}\BibitemShut {NoStop}%
	\bibitem [{\citenamefont {Ciccarello}\ \emph {et~al.}(2022)\citenamefont
		{Ciccarello}, \citenamefont {Lorenzo}, \citenamefont {Giovannetti},\ and\
		\citenamefont {Palma}}]{ciccarello2022quantum}%
	\BibitemOpen
	\bibfield  {author} {\bibinfo {author} {\bibfnamefont {F.}~\bibnamefont
			{Ciccarello}}, \bibinfo {author} {\bibfnamefont {S.}~\bibnamefont {Lorenzo}},
		\bibinfo {author} {\bibfnamefont {V.}~\bibnamefont {Giovannetti}},\ and\
		\bibinfo {author} {\bibfnamefont {G.~M.}\ \bibnamefont {Palma}},\ }\bibfield
	{title} {\bibinfo {title} {Quantum collision models: Open system dynamics
			from repeated interactions},\ }\href
	{https://doi.org/https://doi.org/10.1016/j.physrep.2022.01.001} {\bibfield
		{journal} {\bibinfo  {journal} {Physics Reports}\ }\textbf {\bibinfo {volume}
			{954}},\ \bibinfo {pages} {1} (\bibinfo {year} {2022})}\BibitemShut {NoStop}%
	\bibitem [{\citenamefont {Campbell}\ and\ \citenamefont
		{Vacchini}(2021)}]{campbell2021collision}%
	\BibitemOpen
	\bibfield  {author} {\bibinfo {author} {\bibfnamefont {S.}~\bibnamefont
			{Campbell}}\ and\ \bibinfo {author} {\bibfnamefont {B.}~\bibnamefont
			{Vacchini}},\ }\bibfield  {title} {\bibinfo {title} {Collision models in open
			system dynamics: A versatile tool for deeper insights?},\ }\href
	{https://doi.org/https://doi.org/10.1209/0295-5075/133/60001} {\bibfield
		{journal} {\bibinfo  {journal} {Europhysics Letters}\ }\textbf {\bibinfo
			{volume} {133}},\ \bibinfo {pages} {60001} (\bibinfo {year}
		{2021})}\BibitemShut {NoStop}%
	\bibitem [{\citenamefont {Cattaneo}\ \emph {et~al.}(2021)\citenamefont
		{Cattaneo}, \citenamefont {De~Chiara}, \citenamefont {Maniscalco},
		\citenamefont {Zambrini},\ and\ \citenamefont
		{Giorgi}}]{cattaneo2021collision}%
	\BibitemOpen
	\bibfield  {author} {\bibinfo {author} {\bibfnamefont {M.}~\bibnamefont
			{Cattaneo}}, \bibinfo {author} {\bibfnamefont {G.}~\bibnamefont {De~Chiara}},
		\bibinfo {author} {\bibfnamefont {S.}~\bibnamefont {Maniscalco}}, \bibinfo
		{author} {\bibfnamefont {R.}~\bibnamefont {Zambrini}},\ and\ \bibinfo
		{author} {\bibfnamefont {G.~L.}\ \bibnamefont {Giorgi}},\ }\bibfield  {title}
	{\bibinfo {title} {Collision models can efficiently simulate any multipartite
			markovian quantum dynamics},\ }\href
	{https://doi.org/https://doi.org/10.1103/PhysRevLetters126.130403} {\bibfield
		{journal} {\bibinfo  {journal} {Physical Review Letters}\ }\textbf {\bibinfo
			{volume} {126}},\ \bibinfo {pages} {130403} (\bibinfo {year}
		{2021})}\BibitemShut {NoStop}%
	\bibitem [{\citenamefont {Erbanni}\ \emph {et~al.}(2023)\citenamefont
		{Erbanni}, \citenamefont {Xu}, \citenamefont {Demarie},\ and\ \citenamefont
		{Poletti}}]{erbanni2023simulating}%
	\BibitemOpen
	\bibfield  {author} {\bibinfo {author} {\bibfnamefont {R.}~\bibnamefont
			{Erbanni}}, \bibinfo {author} {\bibfnamefont {X.}~\bibnamefont {Xu}},
		\bibinfo {author} {\bibfnamefont {T.~F.}\ \bibnamefont {Demarie}},\ and\
		\bibinfo {author} {\bibfnamefont {D.}~\bibnamefont {Poletti}},\ }\bibfield
	{title} {\bibinfo {title} {Simulating quantum transport via collisional
			models on a digital quantum computer},\ }\href
	{https://doi.org/https://doi.org/10.1103/PhysRevA.108.032619} {\bibfield
		{journal} {\bibinfo  {journal} {Physical Review A}\ }\textbf {\bibinfo
			{volume} {108}},\ \bibinfo {pages} {032619} (\bibinfo {year}
		{2023})}\BibitemShut {NoStop}%
	\bibitem [{\citenamefont {Garg}\ \emph {et~al.}(2025)\citenamefont {Garg},
		\citenamefont {Ahmed}, \citenamefont {Mitra},\ and\ \citenamefont
		{Chakraborty}}]{garg2025simulating}%
	\BibitemOpen
	\bibfield  {author} {\bibinfo {author} {\bibfnamefont {K.}~\bibnamefont
			{Garg}}, \bibinfo {author} {\bibfnamefont {Z.}~\bibnamefont {Ahmed}},
		\bibinfo {author} {\bibfnamefont {S.}~\bibnamefont {Mitra}},\ and\ \bibinfo
		{author} {\bibfnamefont {S.}~\bibnamefont {Chakraborty}},\ }\bibfield
	{title} {\bibinfo {title} {Simulating quantum collision models with
			hamiltonian simulations using early fault-tolerant quantum computers},\
	}\href {https://doi.org/https://doi.org/10.1103/3trk-smbh} {\bibfield
		{journal} {\bibinfo  {journal} {Physical Review A}\ }\textbf {\bibinfo
			{volume} {112}},\ \bibinfo {pages} {022425} (\bibinfo {year}
		{2025})}\BibitemShut {NoStop}%
	\bibitem [{\citenamefont {Cattaneo}\ \emph {et~al.}(2023)\citenamefont
		{Cattaneo}, \citenamefont {Rossi}, \citenamefont {Garc{\'\i}a-P{\'e}rez},
		\citenamefont {Zambrini},\ and\ \citenamefont
		{Maniscalco}}]{cattaneo2023quantum}%
	\BibitemOpen
	\bibfield  {author} {\bibinfo {author} {\bibfnamefont {M.}~\bibnamefont
			{Cattaneo}}, \bibinfo {author} {\bibfnamefont {M.~A.}\ \bibnamefont {Rossi}},
		\bibinfo {author} {\bibfnamefont {G.}~\bibnamefont {Garc{\'\i}a-P{\'e}rez}},
		\bibinfo {author} {\bibfnamefont {R.}~\bibnamefont {Zambrini}},\ and\
		\bibinfo {author} {\bibfnamefont {S.}~\bibnamefont {Maniscalco}},\ }\bibfield
	{title} {\bibinfo {title} {Quantum simulation of dissipative collective
			effects on noisy quantum computers},\ }\href
	{https://doi.org/https://doi.org/10.1103/PRXQuantum.4.010324} {\bibfield
		{journal} {\bibinfo  {journal} {PRX Quantum}\ }\textbf {\bibinfo {volume}
			{4}},\ \bibinfo {pages} {010324} (\bibinfo {year} {2023})}\BibitemShut
	{NoStop}%
	\bibitem [{\citenamefont {{\c{C}}akmak}\ \emph {et~al.}(2019)\citenamefont
		{{\c{C}}akmak}, \citenamefont {Campbell}, \citenamefont {Vacchini},
		\citenamefont {M{\"u}stecapl{\i}o{\u{g}}lu},\ and\ \citenamefont
		{Paternostro}}]{ccakmak2019robust}%
	\BibitemOpen
	\bibfield  {author} {\bibinfo {author} {\bibfnamefont {B.}~\bibnamefont
			{{\c{C}}akmak}}, \bibinfo {author} {\bibfnamefont {S.}~\bibnamefont
			{Campbell}}, \bibinfo {author} {\bibfnamefont {B.}~\bibnamefont {Vacchini}},
		\bibinfo {author} {\bibfnamefont {{\"O}.~E.}\ \bibnamefont
			{M{\"u}stecapl{\i}o{\u{g}}lu}},\ and\ \bibinfo {author} {\bibfnamefont
			{M.}~\bibnamefont {Paternostro}},\ }\bibfield  {title} {\bibinfo {title}
		{Robust multipartite entanglement generation via a collision model},\ }\href
	{https://doi.org/https://doi.org/10.1103/PhysRevA.99.012319} {\bibfield
		{journal} {\bibinfo  {journal} {Physical Review A}\ }\textbf {\bibinfo
			{volume} {99}},\ \bibinfo {pages} {012319} (\bibinfo {year}
		{2019})}\BibitemShut {NoStop}%
	\bibitem [{\citenamefont {Vu}\ \emph {et~al.}(2026)\citenamefont {Vu},
		\citenamefont {Nguyen}, \citenamefont {M{\"u}stecapl{\i}o{\u{g}}lu},\ and\
		\citenamefont {Ozaydin}}]{vu2026intelligent}%
	\BibitemOpen
	\bibfield  {author} {\bibinfo {author} {\bibfnamefont {D.-K.}\ \bibnamefont
			{Vu}}, \bibinfo {author} {\bibfnamefont {M.~T.}\ \bibnamefont {Nguyen}},
		\bibinfo {author} {\bibfnamefont {{\"O}.~E.}\ \bibnamefont
			{M{\"u}stecapl{\i}o{\u{g}}lu}},\ and\ \bibinfo {author} {\bibfnamefont
			{F.}~\bibnamefont {Ozaydin}},\ }\bibfield  {title} {\bibinfo {title}
		{Intelligent control of collisional architectures for deterministic
			multipartite state engineering},\ }\bibfield  {journal} {\bibinfo  {journal}
		{arXiv preprint arXiv:2602.08526}\ }\href
	{https://doi.org/https://doi.org/10.48550/arXiv.2602.08526}
	{https://doi.org/10.48550/arXiv.2602.08526} (\bibinfo {year}
	{2026})\BibitemShut {NoStop}%
	\bibitem [{\citenamefont {Vidal}\ and\ \citenamefont
		{Dawson}(2004)}]{vidal2004universal}%
	\BibitemOpen
	\bibfield  {author} {\bibinfo {author} {\bibfnamefont {G.}~\bibnamefont
			{Vidal}}\ and\ \bibinfo {author} {\bibfnamefont {C.~M.}\ \bibnamefont
			{Dawson}},\ }\bibfield  {title} {\bibinfo {title} {Universal quantum circuit
			for two-qubit transformations with three controlled-not gates},\ }\href
	{https://doi.org/10.1103/PhysRevA.69.010301} {\bibfield  {journal} {\bibinfo
			{journal} {Physical Review A}\ }\textbf {\bibinfo {volume} {69}},\ \bibinfo
		{pages} {010301} (\bibinfo {year} {2004})}\BibitemShut {NoStop}%
	\bibitem [{\citenamefont {Vatan}\ and\ \citenamefont
		{Williams}(2004)}]{vatan2004optimal}%
	\BibitemOpen
	\bibfield  {author} {\bibinfo {author} {\bibfnamefont {F.}~\bibnamefont
			{Vatan}}\ and\ \bibinfo {author} {\bibfnamefont {C.}~\bibnamefont
			{Williams}},\ }\bibfield  {title} {\bibinfo {title} {Optimal quantum circuits
			for general two-qubit gates},\ }\href
	{https://doi.org/10.1103/PhysRevA.69.032315} {\bibfield  {journal} {\bibinfo
			{journal} {Physical Review A}\ }\textbf {\bibinfo {volume} {69}},\ \bibinfo
		{pages} {032315} (\bibinfo {year} {2004})}\BibitemShut {NoStop}%
	\bibitem [{\citenamefont {Byrd}\ \emph {et~al.}(1995)\citenamefont {Byrd},
		\citenamefont {Lu}, \citenamefont {Nocedal},\ and\ \citenamefont
		{Zhu}}]{byrd1995limited}%
	\BibitemOpen
	\bibfield  {author} {\bibinfo {author} {\bibfnamefont {R.~H.}\ \bibnamefont
			{Byrd}}, \bibinfo {author} {\bibfnamefont {P.}~\bibnamefont {Lu}}, \bibinfo
		{author} {\bibfnamefont {J.}~\bibnamefont {Nocedal}},\ and\ \bibinfo {author}
		{\bibfnamefont {C.}~\bibnamefont {Zhu}},\ }\bibfield  {title} {\bibinfo
		{title} {A limited memory algorithm for bound constrained optimization},\
	}\href {https://doi.org/https://doi.org/10.1137/0916069} {\bibfield
		{journal} {\bibinfo  {journal} {SIAM Journal on Scientific Computing}\
		}\textbf {\bibinfo {volume} {16}},\ \bibinfo {pages} {1190} (\bibinfo {year}
		{1995})}\BibitemShut {NoStop}%
	\bibitem [{\citenamefont {Kingma}\ and\ \citenamefont
		{Ba}(2014)}]{kingma2014adam}%
	\BibitemOpen
	\bibfield  {author} {\bibinfo {author} {\bibfnamefont {D.~P.}\ \bibnamefont
			{Kingma}}\ and\ \bibinfo {author} {\bibfnamefont {J.}~\bibnamefont {Ba}},\
	}\bibfield  {title} {\bibinfo {title} {Adam: A method for stochastic
			optimization},\ }\bibfield  {journal} {\bibinfo  {journal} {arXiv preprint
			arXiv:1412.6980}\ }\href {https://doi.org/10.48550/arXiv.1412.6980}
	{10.48550/arXiv.1412.6980} (\bibinfo {year} {2014})\BibitemShut {NoStop}%
	\bibitem [{\citenamefont {T{\'o}th}\ \emph {et~al.}(2010)\citenamefont
		{T{\'o}th}, \citenamefont {Wieczorek}, \citenamefont {Gross}, \citenamefont
		{Krischek}, \citenamefont {Schwemmer},\ and\ \citenamefont
		{Weinfurter}}]{toth2010permutationally}%
	\BibitemOpen
	\bibfield  {author} {\bibinfo {author} {\bibfnamefont {G.}~\bibnamefont
			{T{\'o}th}}, \bibinfo {author} {\bibfnamefont {W.}~\bibnamefont {Wieczorek}},
		\bibinfo {author} {\bibfnamefont {D.}~\bibnamefont {Gross}}, \bibinfo
		{author} {\bibfnamefont {R.}~\bibnamefont {Krischek}}, \bibinfo {author}
		{\bibfnamefont {C.}~\bibnamefont {Schwemmer}},\ and\ \bibinfo {author}
		{\bibfnamefont {H.}~\bibnamefont {Weinfurter}},\ }\bibfield  {title}
	{\bibinfo {title} {Permutationally invariant quantum tomography},\ }\href
	{https://doi.org/10.1103/PhysRevLett.105.250403} {\bibfield  {journal}
		{\bibinfo  {journal} {Physical Review Letters}\ }\textbf {\bibinfo {volume}
			{105}},\ \bibinfo {pages} {250403} (\bibinfo {year} {2010})}\BibitemShut
	{NoStop}%
	\bibitem [{\citenamefont {Flammia}\ and\ \citenamefont
		{Liu}(2011)}]{flammia2011direct}%
	\BibitemOpen
	\bibfield  {author} {\bibinfo {author} {\bibfnamefont {S.~T.}\ \bibnamefont
			{Flammia}}\ and\ \bibinfo {author} {\bibfnamefont {Y.-K.}\ \bibnamefont
			{Liu}},\ }\bibfield  {title} {\bibinfo {title} {Direct fidelity estimation
			from few pauli measurements},\ }\href
	{https://doi.org/10.1103/PhysRevLett.106.230501} {\bibfield  {journal}
		{\bibinfo  {journal} {Physical Review Letters}\ }\textbf {\bibinfo {volume}
			{106}},\ \bibinfo {pages} {230501} (\bibinfo {year} {2011})}\BibitemShut
	{NoStop}%
	\bibitem [{\citenamefont {da~Silva}\ \emph {et~al.}(2011)\citenamefont
		{da~Silva}, \citenamefont {Landon-Cardinal},\ and\ \citenamefont
		{Poulin}}]{da2011practical}%
	\BibitemOpen
	\bibfield  {author} {\bibinfo {author} {\bibfnamefont {M.~P.}\ \bibnamefont
			{da~Silva}}, \bibinfo {author} {\bibfnamefont {O.}~\bibnamefont
			{Landon-Cardinal}},\ and\ \bibinfo {author} {\bibfnamefont {D.}~\bibnamefont
			{Poulin}},\ }\bibfield  {title} {\bibinfo {title} {Practical characterization
			of quantum devices without tomography},\ }\href
	{https://doi.org/10.1103/PhysRevLett.107.210404} {\bibfield  {journal}
		{\bibinfo  {journal} {Physical Review Letters}\ }\textbf {\bibinfo {volume}
			{107}},\ \bibinfo {pages} {210404} (\bibinfo {year} {2011})}\BibitemShut
	{NoStop}%
\end{thebibliography}

	
\end{document}